\documentclass[10pt,twocolumn,letterpaper]{article}

\usepackage{iccv}
\usepackage{times}
\usepackage{epsfig}
\usepackage{graphicx}
\usepackage{amsmath}
\usepackage{amssymb}

\usepackage{circledsteps}
\pgfkeys{/csteps/fill color=black}
\pgfkeys{/csteps/inner color=white}

\usepackage{amsmath,amsfonts,amsthm,amssymb}
\usepackage{pdfpages}

\usepackage{booktabs}
\usepackage{multirow}
\usepackage{subfigure}
\usepackage{url}
\usepackage{graphicx}
\usepackage{textcomp}
\usepackage{xcolor}
\usepackage{float}
\usepackage{caption}
\usepackage{bbding}
\usepackage{cite}
\usepackage{url}
\usepackage{epstopdf}
\usepackage{enumitem}

\usepackage{algorithmic}
\usepackage[linesnumbered,ruled, vlined]{algorithm2e}

\usepackage[pagebackref=true,breaklinks=true,letterpaper=true,colorlinks,bookmarks=false]{hyperref}

\iccvfinalcopy 

\def\iccvPaperID{****} 

\ificcvfinal\fi

\begin{document}

\title{Ownership Protection of Generative Adversarial Networks}

\author{Hailong Hu\\ SnT, University of Luxembourg
\and
Jun Pang\\
FSTM \& SnT, University of Luxembourg\\
}

\maketitle
\ificcvfinal\thispagestyle{empty}\fi
\thispagestyle{plain}
\pagestyle{plain}

\begin{abstract}
Generative adversarial networks (GANs) have shown remarkable success in image synthesis, making GAN models themselves commercially valuable to legitimate model owners. Therefore, it is critical to technically protect the intellectual property of GANs. Prior works need to tamper with the training set or training process, and they are not robust to emerging model extraction attacks. In this paper, we propose a new ownership protection method based on the common characteristics of a target model and its stolen models. Our method can be directly applicable to all well-trained GANs as it does not require retraining target models. Extensive experimental results show that our new method can achieve the best protection performance, compared to the state-of-the-art methods. Finally, we demonstrate the effectiveness of our method with respect to the number of generations of model extraction attacks, the number of generated samples,  different datasets, as well as adaptive attacks.
\end{abstract}

\section{Introduction}
\label{sec:intro}

Generative adversarial networks~(GANs), as one of the most successful generative models, have already exerted revolutionary influences on many application domains, such as image synthesis~\cite{goodfellow2014generative, SNGAN2018spectral, StyleGAN12019style, BigGAN2018large,sauer2022stylegan}, image editing~\cite{zhu2020domain,shen2020closed, cherepkov2021navigating,xia2022gan}, and image translation~\cite{zhu2017unpaired, yi2017dualgan,ma2018gan,isola2017image}.
However, building a well-trained state-of-the-art GAN model is not straightforward. 
It usually requires the complicated and exhausting process of data collection, expert-level knowledge in model architecture design, elaborate hyperparameter tuning, and extensive computing resources.
Therefore, a high-quality GAN model is incredibly costly and should be regarded as the intellectual property of the model owner.

As GAN models are valuable, this simultaneously incentivizes adversaries to steal these models in various ways.
On the one hand, adversaries can physically steal a GAN model via malware infection or insider attacks~\cite{schultz2002framework}. 
An insider attack can directly copy the victim model through those who are authorized to access the full model.
As a result, the stolen model is totally the same as the victim model.
On the other hand, adversaries can functionally steal a GAN model via model extraction attacks~\cite{hu2021model}. 
This threat exists because an increasing number of technology companies provide Machine Learning as a Service~(MLaaS) to their customers, such as Amazon AWS, Google Cloud, and Microsoft Azure.
A model extraction attack enables adversaries to obtain a substitute model via exposed interfaces.
As a consequence, the stolen model is functionally similar to the victim model.
In general, 
both physical stealing attacks and model extraction attacks seriously jeopardize the intellectual property of legitimate model owners.
It is paramount to develop ownership protection methods to safeguard the intellectual property of a GAN model.

Despite confronting these threats, research on ownership protection for GANs is somehow much less explored.
There is one work~\cite{ong2021protecting} that proposes a watermark-based method for protecting ownership of GANs, but it needs to utilize extra loss functions to retrain the model.
More importantly, it only considers physical stealing, and emerging model extraction attacks are taken into consideration.

In this paper, we develop a new ownership protection method for GANs, which can protect ownership on both physical stealing and model extraction attacks.
Our method claims the ownership of a GAN 
by leveraging common characteristics of a target model and its stolen models.
The rationale for our method is that stolen models are derived from the target model while honest models are not. Thus, these common characteristics can be leveraged to differentiate stolen models from honest models.
More specifically, we utilize generated samples from GANs to build a discriminative classifier to learn these characteristics.
This is because the objective of a GAN is to learn the distribution of a training dataset and the learned implicit distribution of a GAN can be represented by these generated samples~\cite{StatisticalMechanics} (see Section~\ref{ssec:own_pro_alg}).

We comprehensively evaluate our new method by comparing it with two state-of-the-art works: watermark-based method~(abbreviation as Ong method)~\cite{ong2021protecting} and fingerprint-based method~(abbreviation as Yu method)~\cite{yu2021artificial}.
Here, note that fingerprint-based methods are proposed for deepfake detection and attribution~\cite{wang2020cnn,yu2019attributing,yu2021artificial,yu2022responsible,marra2019gans}. In this work, \textit{we are the first to introduce them to the ownership protection field}, considering their common objective: fingerprints can be used to infer whether a suspect sample is from the model.
Extensive experiments show that our method achieves the best performance in all evaluations among these ownership protection methods.
In contrast, the other two methods cannot work in many cases, especially for model extraction attacks~(see Section~\ref{sec:evaluation}).
Furthermore, we analyze the protection performance by visualizing the characteristics learned by our method~(see Section~\ref{ssec:distri}).
We also show the stability of our method on the number of generations of model extraction attacks, which is a new emerging threat in the generative domain.
In contrast, the protection performance of the Yu method presents a significant linear decrease and the Ong method completely fails in providing any protection~(see Section~\ref{ssec:ana_generations}).
Our analysis with respect to the number of generated samples and different datasets further shows the effectiveness of our method.
Finally, extensive evaluations under adaptive attacks also demonstrate that our method is still effective and robust even if adversaries obtain partial knowledge of our method~(see Section~\ref{sec:MPAdA}).

In summary, we make the following contributions.
(1) We propose a novel ownership protection method GAN-Guards from a new perspective: detecting ownership infringement by utilizing the common characteristics of a target model and stolen models.
(2) We evaluate our method on various attacks, including emerging model extraction attacks.
(3) Our method achieves a new state-of-the-art performance in ownership protection for GANs compared to prior works. 
The source code will be made public along with the final version of the paper.

\section{Related Work}
\label{sec:Related work}

\noindent
\textbf{Generative adversarial networks.}
Generative adversarial networks~(GANs) have undoubtedly achieved a series of successes in image generation~\cite{SNGAN2018spectral,PGGAN2018progressive,StyleGAN12019style, StyleGAN22019analyzing,BigGAN2018large}, image manipulation~\cite{shen2020interpreting,shen2020closed}, and image super-resolution~\cite{ledig2017photo,zhang2019ranksrgan}.
The seminal GAN introduced in 2014~\cite{goodfellow2014generative} presents a promising result in image synthesis, which significantly inspires more and more researchers to propose various methods to further advance the performance of GANs.
Karras et al.~\cite{PGGAN2018progressive} propose a progressive training strategy that enables a GAN to synthesize high-resolution images.
In addition to generating high-resolution images, Karras et al.~\cite{StyleGAN12019style} further introduce neural style transfer structures into the architecture of GANs and it empowers GANs to generate a variety of style images.
Takeru et al.~\cite{SNGAN2018spectral} introduce spectral normalization to normalize the weights of each layer to improve the quality of synthetic images. 
\textit{Overall, each improvement in GANs requires talented researchers to devote tremendous efforts.
In this work, instead of further improving the performance of GANs, we target at developing a technique to protect the intellectual property of valuable GANs.}

\smallskip\noindent
\textbf{Ownership protection.}
There are numerous works aiming to protect ownership of discriminative models~\cite{uchida2017embedding,zhang2018protecting, jia2021entangled, chen2021copy,lukasdeep,li2019prove,dziedzicdataset,cong2022sslguard}.
These works can be generally classified into three groups: 
embedding watermarks into model parameters~\cite{uchida2017embedding}, using predefined inputs as triggers~\cite{adi2018turning} and utilizing unique features of models~\cite{maini2020dataset,chen2021copy}.
\textit{However, methods on discriminative models cannot be applied to generative models since they are different machine learning models.}

There are only a few works on ownership protection of generative models.
Ong et al.~\cite{ong2021protecting} propose a protection framework for GANs by adding a novel regularization term to the existing loss function.
Some works on fingerprints of GANs can be applied into this field.
Yu et al.~\cite{yu2021artificial} propose to add fingerprints into training data and then verify the fingerprints on GANs.
Additionally, Yu et al.~\cite{yu2022responsible} add a novel fingerprint embedding layer to modulate the generation of fingerprinted images.
\textit{However, these works do not perform evaluations on emerging model extraction attacks. In this work, we propose a novel protection method and thoroughly evaluate these works on various attacks, including model extraction attacks.}

\section{Background}
\label{sec:Background}

\subsection{Generative Adversarial Networks}
An unconditional GAN generally consists of a generator~$G$ and a discriminator~$D$. 
In the training phase, the generator~$G$ aims to generate fake data to fool the discriminator~$D$ while the discriminator~$D$ attempts to distinguish fake data from the generator from the real data from the training set. 
Once finishing training, the generator~$G$ can be utilized to generate data, given latent codes.
Gaussian distribution or uniform distribution is commonly used to obtain latent codes.
Mathematically, the generator of a GAN is a function $G: \mathcal Z \rightarrow \mathcal X$ that maps a low dimensional latent space $\mathcal Z \subseteq \mathbb R^n$ to a high dimension data space $\mathcal X \subseteq \mathbb R^m$.

\subsection{Paradigms of Ownership Protection}
\label{ssec:para_own_pro}
Current paradigms of ownership protection on GANs can be divided into two classes: watermark-based methods and fingerprint-based methods.
Watermark-based methods initially are proposed in the work~\cite{ong2021protecting}. 
The key idea is that model owners claim ownership of this GAN if special outputs can be obtained from a suspect GAN.
Generally, these special outputs (e.g. watermarked generated samples) are obtained through special queries from a GAN.
For simplicity, we refer to this ownership protection method by the first author's name, i.e, Ong~\cite{ong2021protecting}.

Fingerprint-based methods are initially proposed for deepfake detection and attribution~\cite{yu2021artificial,yu2022responsible}. 
In this work, we apply these methods to protect ownership of GANs.
This is because they share a common objective that fingerprints can be used to infer whether a suspect sample is from their model. 
Specifically, the key idea of fingerprint-based methods is that if the fingerprint extracted from generated samples from a GAN is identical to the true fingerprint, model owners can claim ownership of the GAN.
To achieve this goal, various methods are proposed, such as adding fingerprints on the training set of a GAN~\cite{yu2021artificial}, and designing new architectures of a GAN and loss functions~\cite{yu2022responsible}.
In this work, considering the flexibility and scalability of a method, we choose the former method---Yu~\cite{yu2021artificial} to evaluate the performance in ownership protection and make a comparison with our proposed method.

Note that, unlike these paradigms that require forcibly implanting watermarks/fingerprints into target models and retraining target models, our method provides a novel paradigm: the common characteristics of a target model and its stolen models are exploited to claim ownership, motivated by the emerging of model extraction attacks.

\section{A New Ownership Protection Method}
\label{sec:Methods}

\subsection{Threat Model}

We assume that defenders, i.e. model providers who deploy an ownership protection method on their target model, only have access to generated samples from a suspect model deployed by the adversaries. 
Thus, the defenders make an ownership infringement decision only based on these generated samples.
\textit{This is the most practical and strictest assumption for defenders.}

\subsection{Key Observations}

The first key observation is that physical stealing and model extraction attacks are two fundamental but different types of ownership infringement.
Physical stealing attacks refer that an adversary physically copies a model~$G_{\it sub}$ from the target model~$G_{\it tar}$. Therefore, $G_{\it sub}$ is totally the same as $G_{\it tar}$.
Model extraction attacks~\cite{hu2021model} refer that an adversary retrains a substitute model~$G_{\it sub}$ on generated samples from a target model $G_{\it tar}$.
These generated samples can be obtained by an adversary when model owners release generated samples or provide a querying interface.
Thus, $G_{\it sub}$ is functionally similar to $G_{\it tar}$.

The second key observation is that as stolen models~(i.e. constructed by physical stealing or model extraction attacks) are derived from the target model but honest models are not, it is thus natural to assume that stolen models and the target model share common characteristics which do not exist in honest models. 
Therefore, we can learn and leverage such characteristics as an evidence to differentiate stolen models from honest models.

\begin{figure}[!t]
	\centering
	\includegraphics[width=0.87\linewidth]{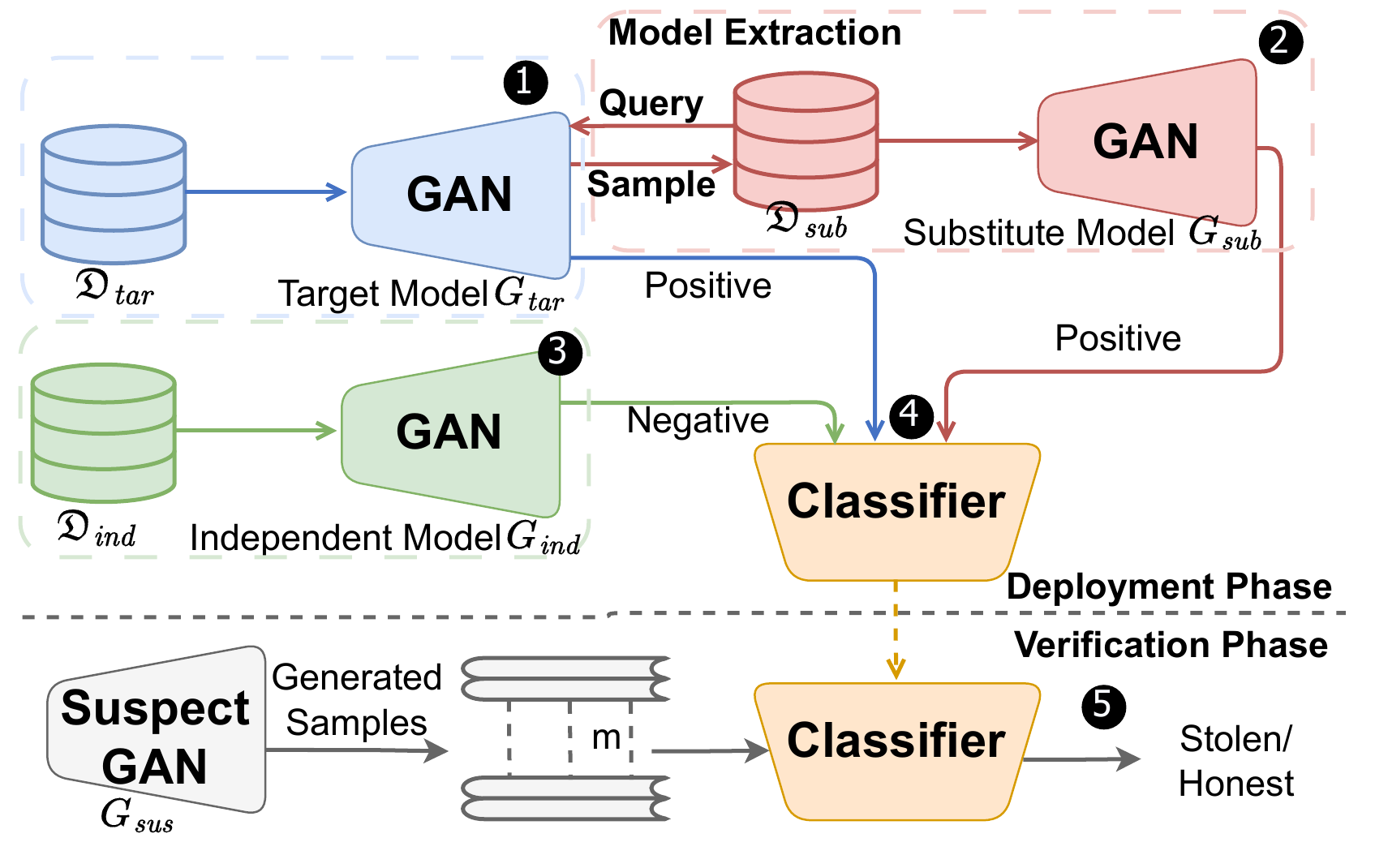}
	\caption{Overview of our method.   \Circled{\small\textbf{1}}~A target model is trained on a dataset~$\mathfrak D_{\it tar}$. \Circled{\small\textbf{2}}~A substitute model is constructed by model extraction. \Circled{\small\textbf{3}}~An independent model is trained on a dataset~$\mathfrak D_{\it ind}$ that has the same distribution as the dataset~$\mathfrak D_{\it tar}$, but it is disjoint with $\mathfrak D_{\it tar}$.  \Circled{\small\textbf{4}} A classifier is trained to discriminate between stolen models and honest models. \Circled{\small\textbf{5}} The classifier is used for the verification of a suspect model.	
	Here, the target model can also refer to the physically stealing model. The defenders do not have any information about a suspect model~$G_{\it sus}$, except generated samples, in the verification phase.}
	
	\label{fig:ow_method}
\end{figure} 

\subsection{Ownership Protection Algorithm}
\label{ssec:own_pro_alg}

\noindent
\textbf{Overview.}
In order to extract the characteristics for a target model, our method proposes to learn these characteristics by training a binary classifier on generated samples. 
Generated samples from model extraction and physical stealing are labelled as positive while samples from honest models, i.e. independently trained models,  are labelled as negative. The reason why we use generated samples is that a GAN model is to learn the distribution of a training set. The learned distribution is implicit, which can be represented through these generated samples \cite{StatisticalMechanics}. 
This process is shown in the deployment phase of Figure~\ref{fig:ow_method}.

In practice, it is impossible for the defenders to consider all independently trained models and all models constructed by model extraction.
Therefore, our method constructs positive and negative GAN models only by the limited knowledge of the defenders: the architectures of the target model and the training set.  
Specifically, the architectures of all models (i.e. $G_{\it tar}$, $G_{\it sub}$, $G_{\it ind}$) in the deployment phase are the same. 
The independent set~$\mathfrak D_{\it ind}$ and the target set~~$\mathfrak D_{\it tar}$ are from the same distribution but disjoint.
$\mathfrak D_{\it ind}$ can be easily obtained, e.g. the defenders split a dataset into two parts: one as the independent set and the other as the training set.  
We emphasize that our method is practical as suspect models constructed by the adversaries can be trained on any (unknown for the defenders) GAN architectures and datasets. 
Our extensive experiments in Section~\ref{ssec:Gen_ME}, demonstrate our method can well generalize beyond these unknown GANs and correctly recognize them.

\begin{algorithm}[!t]
	\caption{The GAN-Guards Algorithm.}
	\label{alg:gan-gards}	
	\small
	\KwIn{a target model: $G_{\it tar}$; 
		an independent dataset $\mathfrak D_{\it ind}$;
		$m$ samples $X_{\it sus}$ from a suspect model $G_{\it sus}$.}
	\KwOut{ownership decision: ${\it OwDecision}$}
	\vspace{1mm}
	\SetKwFunction{FSub}{buildProtection}
	\SetKwProg{Def}{def}{:}{}
	\Def{\FSub{$G_{\it tar}, \mathfrak D_{\it ind}$ }}{
		
		Sample $\tilde n$ samples $X_{\it gen}$ from $G_{\it tar}$\;
		$G_{\it sub} \leftarrow {\sf trainGAN}(X_{\it gen})$\;			
		$G_{\it ind} \leftarrow {\sf trainGAN}(\mathfrak D_{\it ind})$\;		
		
		Sample $n$ samples $X_{\it gen}$ from $G_{\it tar}$;   \textcolor{gray}{$\triangleright$  labelling positive for physical stealing.}\
		
		Sample $n$ samples $X_{\it sub}$ from $G_{\it sub}$; \textcolor{gray}{$\triangleright$  labelling positive for model extraction.}\
		
		Sample $2n$ samples $X_{\it ind}$ from $G_{\it ind}$; \textcolor{gray}{$\triangleright$  labelling negative for the honest model.}\
		
		${\it Classifier} \leftarrow {\sf trainClassifier}(X_{\it gen}, X_{\it sub}, X_{\it ind})$\;			
		
		\Return ${\it Classifier}$
	}
	
	\vspace{1mm}
	\SetKwFunction{FSub}{performVerification}
	\SetKwProg{Def}{def}{:}{}
	\Def{\FSub{${\it Classifier}, X_{\it sus}, \tau$ }}{
		Initialize prediction array $\it pred$ of length $m$ with $0$\;	
		\For{$i=0$ \KwTo $m-1$}{	
			
			${{\it pred}[i]} \leftarrow {\it Classifier}(X_{\it sus}[i])$;  \textcolor{gray}{$\triangleright$  Prediction: 1 or 0.}\
		}
		${\it ConfiSocre} = {\sf sum({\it pred})}/{m}$ \;
		\textcolor{gray}{$\triangleright$  Make a decision based on multiple samples.}\
		
		\lIf{${\it ConfiSocre} > \tau$} {
			${\it OwDecision} = 1$}	
		\lElse {${\it OwDecision} = 0 $} 
		
		\Return ${\it OwDecision}$
	}	
	
\end{algorithm}

\smallskip\noindent
\textbf{Algorithm.}
As illustrated in the function {\it buildProtection} of Algorithm~\ref{alg:gan-gards},
specifically, given the generator of a target model~$G_{\it tar}$ and an independent dataset $\mathfrak D_{\it ind}$, we first construct a substitute model~$G_{\it sub}$ by extracting the target model~$G_{\it tar}$. 
Next, we train a GAN~$G_{\it ind}$ on the independent dataset $\mathfrak D_{\it ind}$.
Samples from~$G_{\it sub}$ and $G_{\it tar}$ are labeled as positive while samples from~$G_{\it ind}$ are labeled as negative.
These samples are used for training a classifier and in this work we choose ResNet50~\cite{he2016deep} as the classifier. 

After obtaining the trained classifier, we start to perform the verification of ownership. 
We first collect $m$ generated samples released by a suspect model~$G_{\it sus}$. 
These samples are fed into the classifier and $m$ predictions can be obtained. 
We calculate the percentage of these positive predictions. If it is larger than a predefined threshold, the suspect model is inferred as stealing from the target model.
We also analyze how the number of generated samples~$m$ affects our performance in Section~\ref{ssec:ana_queries}.
This process is also illustrated in the function {\it performVerification} of Algorithm~\ref{alg:gan-gards}.

\section{Experiments}
\label{sec:Experiments}

\subsection{Datasets}
\label{ssec:Datasets}
We evaluate our method on two datasets: FFHQ~\cite{StyleGAN12019style} and Church~\cite{yu2015lsun}. 
They are typically used in image generation.
The FFHQ dataset is designed for human face image synthesis and includes 70,000 images.
The Church dataset is from the LSUN dataset, which contains 126,277 outdoor church images.

All images are resized to $64 \times 64$. 
For each dataset, we randomly split the dataset into three disjoint equal parts and mark each part as the corresponding dataset name plus `I', `II', and `III', respectively, such as FFHQ-I and FFHQ-II.
Dataset I, i.e.~$\mathfrak D_{\it tar}$, is used to train a target GAN model.
Dataset II is used to train a GAN and later the model (i.e. Ind-a in Section~\ref{ssec:susp_models}) is used as negative to test the ownership protection methods. 
Dataset III, i.e.~$\mathfrak D_{\it ind}$, is used to train a GAN model and later the model is used for building a classifier together with the target model.
Specifically, we set the size of each part of FFHQ and Church as 20,000 and 40,000, respectively.

\subsection{Suspect Models}
\label{ssec:susp_models}
We consider various suspect models. Positive suspect models are considered ownership infringement and these models are derived from the target models via physical stealing and model extraction, and obfuscation attacks, such as input perturbation, output perturbation, overwriting, and fine-tuning attacks.
Negative suspect models are host models and they are built from independent training. Here, we consider two types: \textit{Ind-a} trained on dataset II with the same architectures of target models, and \textit{Ind-b} that is trained on dataset I, with the same architectures of target models but uses different seeds, i.e. different random initializations. 
In Section~\ref{ssec:Suspect_Models} in Appendix, we detail the implementation and report performance of these suspect models.

\subsection{Metrics}
We use FID~\cite{FID2017gans} to measure the performance of a GAN. 
50K generated samples from a GAN and all training samples are used to compute the FID value.

In terms of protection performance, the Ong method~\cite{ong2021protecting} utilizes the SSIM~\cite{wang2004image} score to measure the similarity between the groundtruth watermark and the watermark extracted from a suspect model. 
If the SSIM score of an image is higher than a threshold,  the image is more likely from the target model. 
The Yu method~\cite{yu2021artificial} calculates a bitwise accuracy between the groundtruth fingerprint and an extracted fingerprint. 
Claiming ownership of a model based on only one image is not robust enough.
Therefore, we make a final decision by computing a confidence score on multiple samples. 
Specifically, given $m$ samples and each sample gets an output $o \in \{0,1\}$ from a suspect method, the confidence score that recognizes a suspect model as positive is computed by: ${\it Confidence Score = \frac{\sum_{i=0}^{m-1} o_i}m}$.
In this work, we set threshold~$\tau$ of all methods as 90\% for consistency.
Thus, a suspect model is predicted as positive (stolen model) if $\tau \geq 90\%$.
We fix the number of samples~$m$ as 1,000.

\subsection{Experimental Setups}

In terms of GANs, we use SNGAN~\cite{SNGAN2018spectral}, PGGAN~\cite{PGGAN2018progressive} and StyleGAN~\cite{StyleGAN12019style}. 
These all can achieve excellent performance in image synthesis.
We use the official implementation of each GAN to train GANs.
For model extraction attacks, considering a trade-off between attack cost and performance, we set the number of generated samples as 50,000, which is also suggested by the work~\cite{hu2021model}. 

For our protection method, we use ResNet50~\cite{he2016deep} pretrained on ImageNet~\cite{russakovsky2015imagenet} dataset for our classifier. 
The SGD optimizer with a learning rate of 0.003 is used and the number of epochs is fixed as 5.
As shown in Algorithm~\ref{alg:gan-gards}, we use the Gaussian prior distribution to obtain generated samples and the number of samples~$n$ is set as 100,000.
Therefore, 400,000 samples in total are used for training the classifier. 
For the Ong method~\cite{ong2021protecting} and the Yu method~\cite{yu2021artificial}, we adopt their official implementations with suggested hyperparameters.

\section{Evaluation}
\label{sec:evaluation}
In this section, we compare our  method with two state-of-the-art methods: the Ong method~\cite{ong2021protecting}  and the Yu method~\cite{yu2021artificial}, and both have been already discussed in Section~\ref{ssec:para_own_pro}.
We evaluate them from various perspectives, including model utility, verification performance, robustness to obfuscations, and robustness to more model extractions.

\begin{table}[!t]
	\centering
	\caption{Performance of target model SNGAN trained on FFHQ-I on different methods. ($\downarrow$ is better).}
	\label{tab:ModelUtility}
	\renewcommand{\arraystretch}{1.0}
	\scalebox{1}{
\begin{tabular}{lccc} 
\specialrule{.15em}{.05em}{.05em}
	Methods                                & Ong   & Yu    & Ours   \\ 
	\midrule
	FID$(\mathfrak D,\tilde G) \downarrow$ & 20.14 & 26.46 & 20.25  \\

\specialrule{.15em}{.05em}{.05em}
\end{tabular}

}
\end{table}

\subsection{Model Utility}
\label{ssec:Model_Utility}
Table~\ref{tab:ModelUtility} shows the performance of the target model SNGAN trained on FFHQ-I with different protection methods. 
The FID is computed by the original training set~$\mathfrak D$ and the protected GAN~$\tilde G$.
Overall, the watermark-based method Ong and our method achieve similar outstanding performance, while the fingerprint-based method Yu shows worse performance.
This is because the Yu method needs to add fingerprints into a training set, which is at the cost of sacrificing model utility.

\subsection{Verification Performance}
\label{ssec:Eval_BA}

\begin{figure}[!t]
	\centering
	\includegraphics[width=0.7\linewidth]{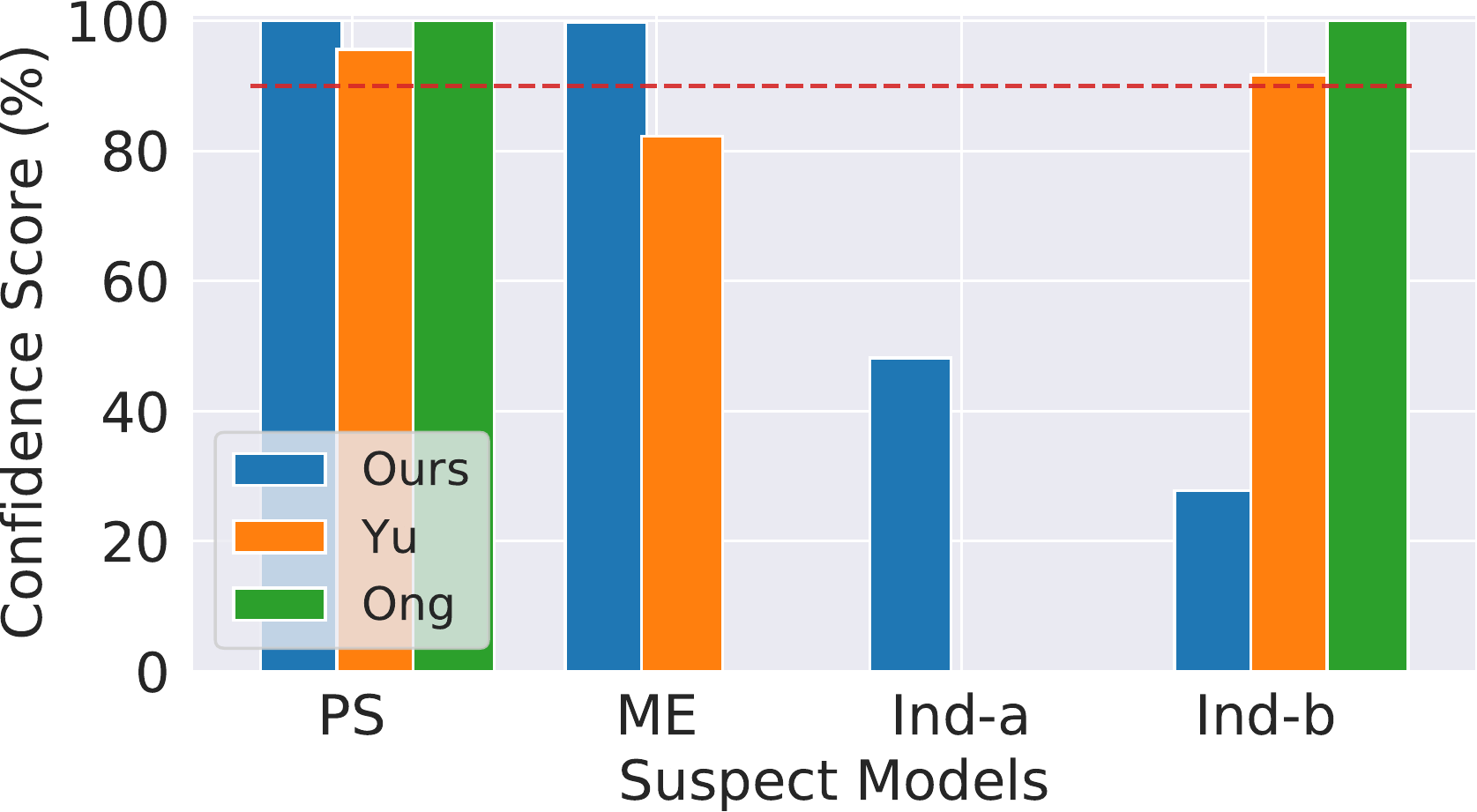}
	\caption{Verification performance of all methods. The target model SNGAN is trained on FFHQ-I. PS and ME are positive suspect models while Ind-a and Ind-b are negative suspect models. Note that \emph{green and orange bars in some cases cannot be observed because their scores are 0\%.} }
	\label{fig:cmp_basic}
\end{figure}

Figure~\ref{fig:cmp_basic} presents verification performance on different ownership protection methods. 
The red dashed line is the threshold~$\tau$ of the confidence score. 
A model is predicted as a stolen (positive) model if $\tau \geq 90\%$.
PS refers to models from physical stealing while ME refers to models from model extraction. 
Note that here ME, Ind-a, Ind-b models used in the verification phase are not the same models used in our deployment phase~(detailed in Section~\ref{ssec:susp_models}).
Overall, our method can correctly differentiate all positive and negative suspect models, achieving 100\% accuracy.
However, the Ong method and the Yu method are unable to defend against model extraction attacks.
Additionally, the Ong and Yu methods mistakenly recognize the suspect model Ind-b trained with different initializations as a stolen model.
This is because embedded watermarks or fingerprints cannot be changed only owing to different initializations of a training process.
Thus, their methods lead to false alarms and hurt honest model providers.

\subsection{Robustness to Obfuscations}
\label{ssec:Eval_AdvA}

In order to evade verification, advanced adversaries may utilize obfuscation techniques to obfuscate stolen models.
In this work, we consider four types of obfuscation techniques: input perturbation, output perturbation, overwriting and fine-tuning. 
Input perturbation aims to modify the queries, i.e., latent codes, to evade special queries.
Here, we adopt random input perturbation. 
That is, for any query, a target model resamples
latent codes from Gaussian distribution.
For brevity, we rename it Inp.
Output perturbation refers to perturbing generated samples by various post-processing techniques.
We use four different output perturbations: additive Gaussian noise, Gaussian filter, Gaussian blurring, and JPEG compression. 
We briefly rename them Oup-a, Oup-b, Oup-c, and Oup-d, respectively. 
The magnitude of these perturbations is set as 0.01, 0.4, 0.5, and 0.85, respectively. 
Overwriting refers to encoding a different watermark/fingerprint to overwrite the original watermark/fingerprint. Our method does not rely on watermarks and fingerprints, thus intrinsically eliminating the threat of this attack.
In this work, we consider wholly fine-tuning where we take the weights of the stolen model as initialization and retrain a GAN model on a different dataset FFHQ-II.
Because these obfuscation operations can be added into physical stealing (PS) or model extraction (ME),  there are different combinations between obfuscation operations and PS and ME.
Here, we mark them as 'PS+` and 'ME+` corresponding obfuscation operations, such as PS+Inp.
Implementations are illustrated in Section~\ref{ssec:details_obfu} in Appendix.

\begin{figure*}[!t]
	\centering
	
	\subfigure[Input perturbation.]{
		\includegraphics[width=0.60\columnwidth]{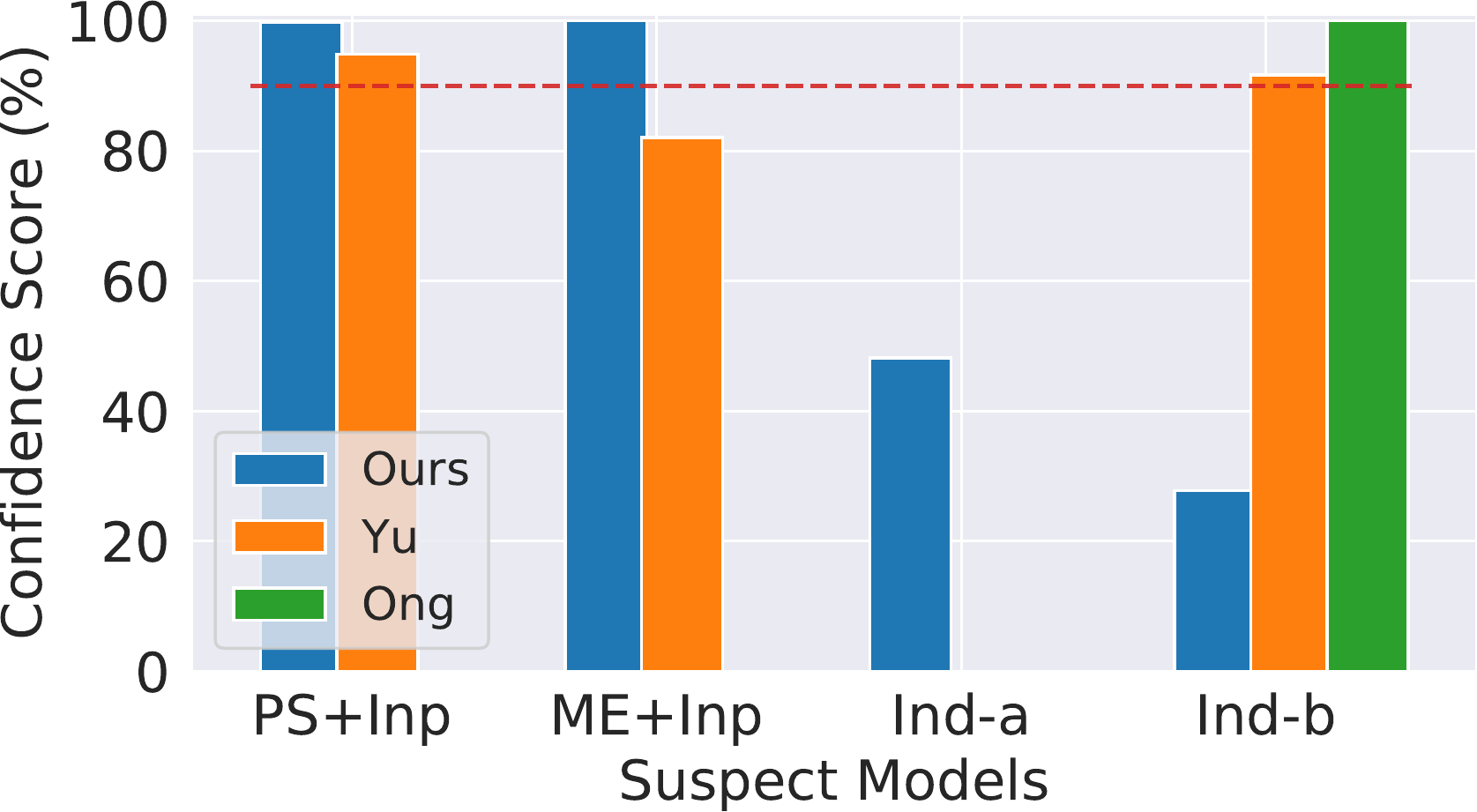}
		\label{fig:cmp_inp}
	}
	\subfigure[PS + output perturbation.]{
		\includegraphics[width=0.60\columnwidth]{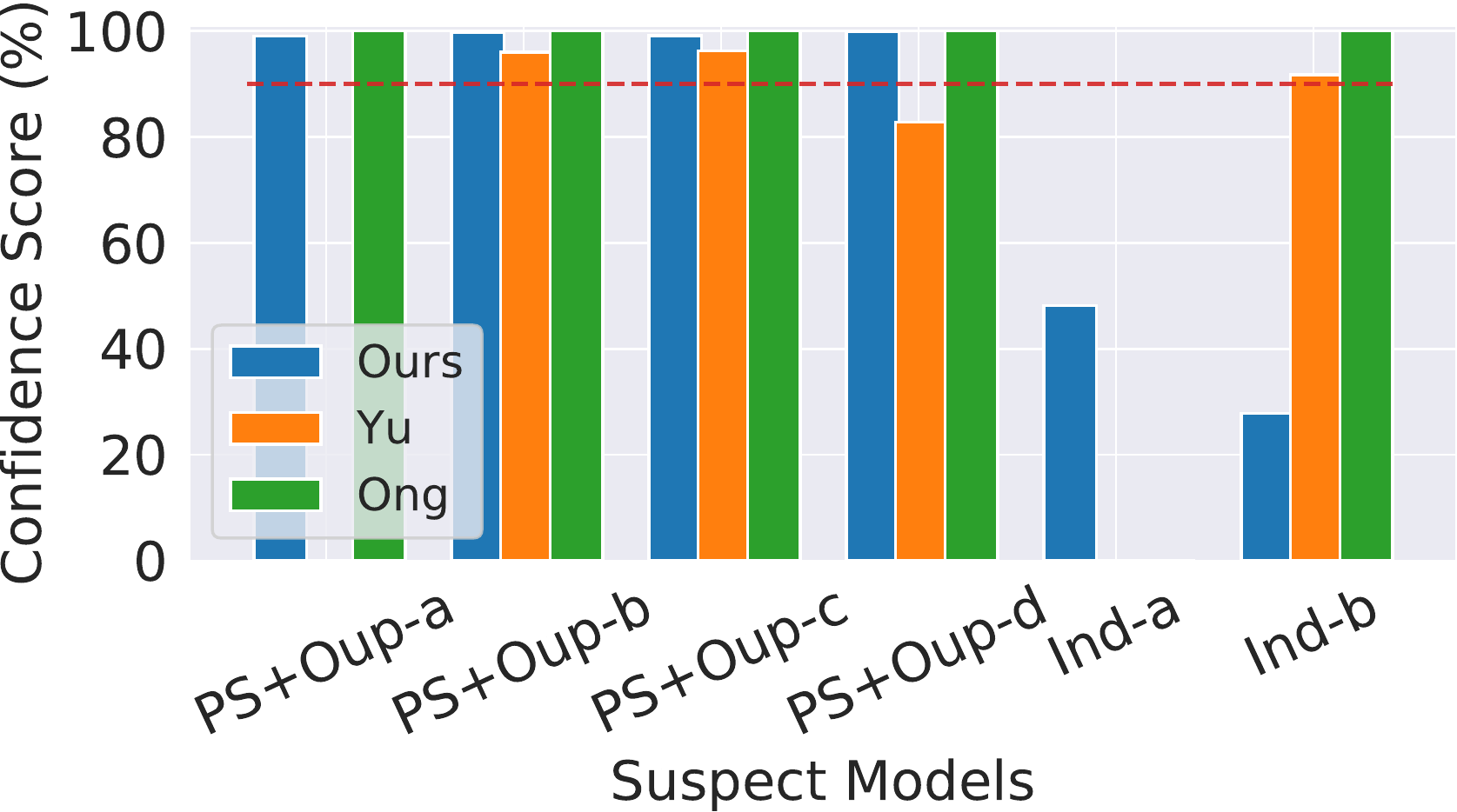}
		\label{fig:cmp_oup_ps}
	}	
	\subfigure[ME + output perturbation.]{
		\includegraphics[width=0.60\columnwidth]{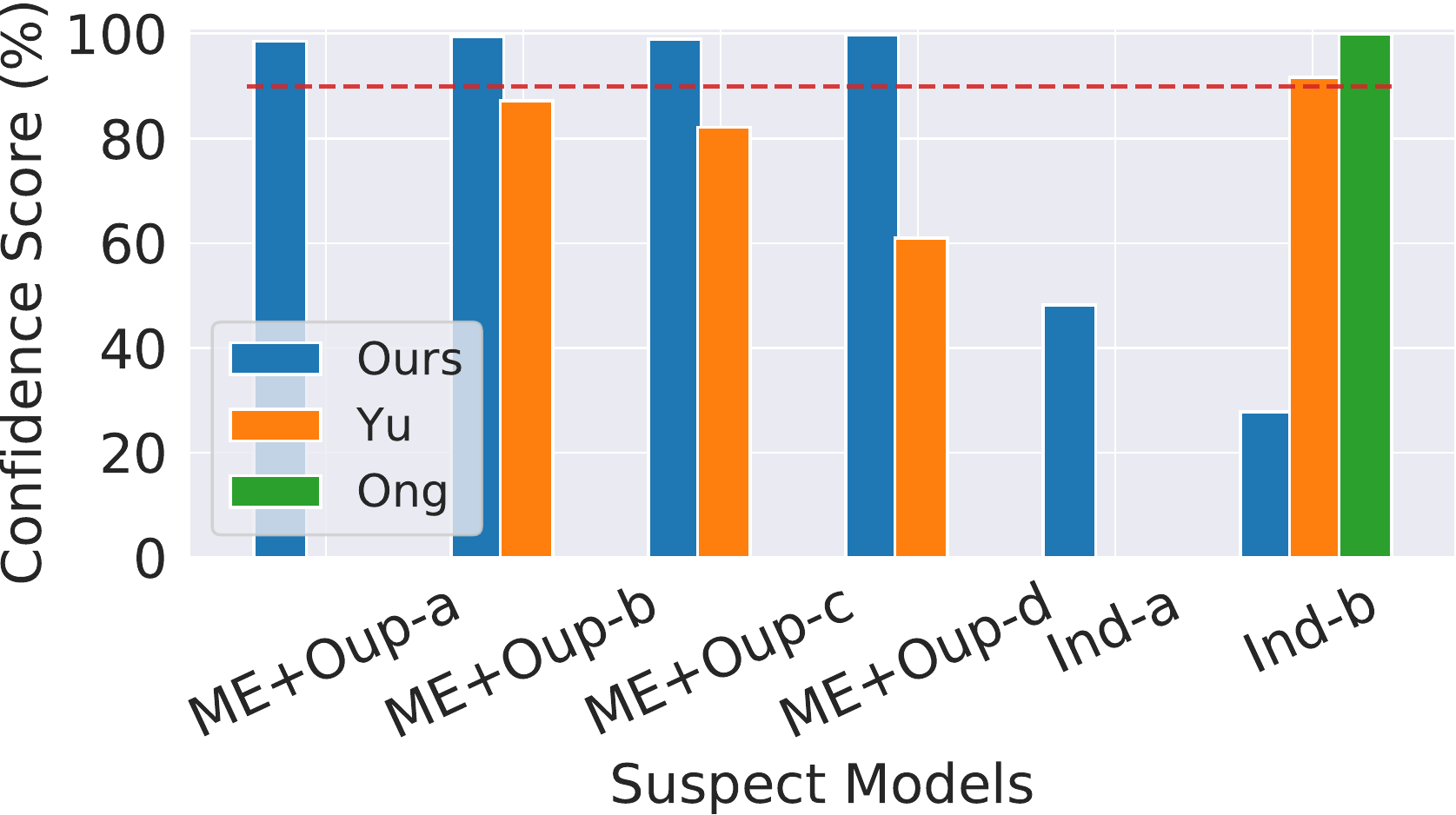}
		\label{fig:cmp_oup_me}
	}
 \vspace{-3mm}
	\caption{Robustness to Obfuscations. Protection performance for target model SNGAN trained on FFHQ-I. Again, green and orange bars in some cases cannot be observed because their scores are 0\% and cannot defend against these attacks.}
	\label{fig:Eval_oba}
\end{figure*} 

\smallskip\noindent
\textbf{Results.}
Figure~\ref{fig:Eval_oba} shows the protection performance under input and output perturbation operations.
Overall, our method can still remain 100\% accuracy against input perturbation and output perturbation attacks.
In contrast, the Ong method totally cannot resist the input perturbation attack, as shown in Figure~\ref{fig:cmp_inp} and the Yu method cannot defend against additive Gaussian noise of output perturbation attacks (PS+Oup-a and ME+Oup-a).
Again, Figure~\ref{fig:cmp_oup_me} shows that the Ong and Yu methods cannot defend against ME+Output perturbation. 
We analyze the reasons for the Ong and Yu methods in Section~\ref{ssec:explan_ong} in Appendix.

We perform the evaluation under the overwriting attack. We do not report results for our method because our method does not rely on watermarks or fingerprints. 
Overall, the Ong and Yu methods cannot defend against this type of attack and both confidence scores are 0\%.
It indicates that the overwritten watermarks and fingerprints make their methods unable to extract the expected outputs.
We summarize the results in Table~\ref{tab:Eval_AdvA_sngan_ffhq1_overwriting} in Appendix. 

We evaluate the protection performance under the fine-tuning attack. Unfortunately, we observe that all methods are not robust to the fine-tuning attack. 
We analyze that this is because the fine-tuned GAN model has learned a different distribution in a new training set and neural networks suffer from catastrophic forgetting~\cite{french1999catastrophic,goodfellow2013empirical}. 
The former makes that our method recognizes this model as an independent training model while the latter makes that the Ong and Yu methods forget embedded watermarks and fingerprints.  
This also inspires us to think about the ownership boundary of a GAN and develop more powerful protection work in future.
We summarize the results in Table~\ref{tab:Eval_AdvA_sngan_ffhq1_finetuning*} in Appendix.

\begin{figure}[!t]
	\centering
	\includegraphics[width=0.68\linewidth]{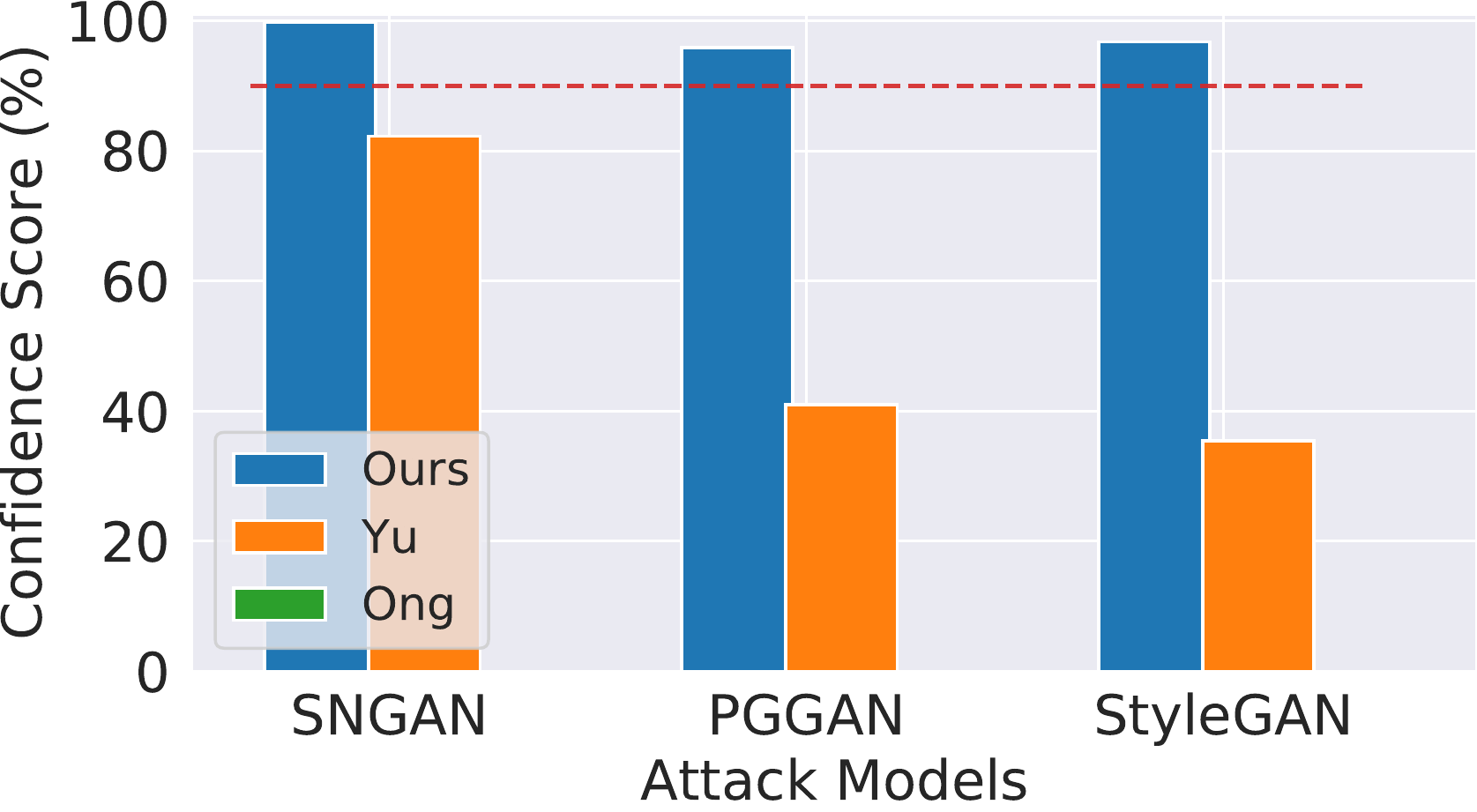}
	\caption{Protection performance under model extraction attacks with different GANs as attack models. The target model SNGAN is trained on FFHQ-I.}
	\label{fig:cmp_More_ME}
\end{figure} 

\subsection{Robustness to More Model Extraction}
\label{ssec:Gen_ME}
When mounting model extraction attacks, adversaries can utilize various architectures of GANs  to extract a target model. 
Figure~\ref{fig:cmp_More_ME} shows the results of all methods in terms of robustness to model extraction attacks with different GANs as attack models. The target model is SNGAN.
We see that our method still performs well, while the other two methods recognize them as honest models. 
\textit{This shows our method can recognize models constructed by model extraction attacks regardless of the GAN architectures of adversaries.}

\section{Analysis}
\label{sec:aba}

\subsection{Visualization of Characteristics}
\label{ssec:distri}

\begin{figure}[!t]
	\centering
	\includegraphics[width=0.8\linewidth]{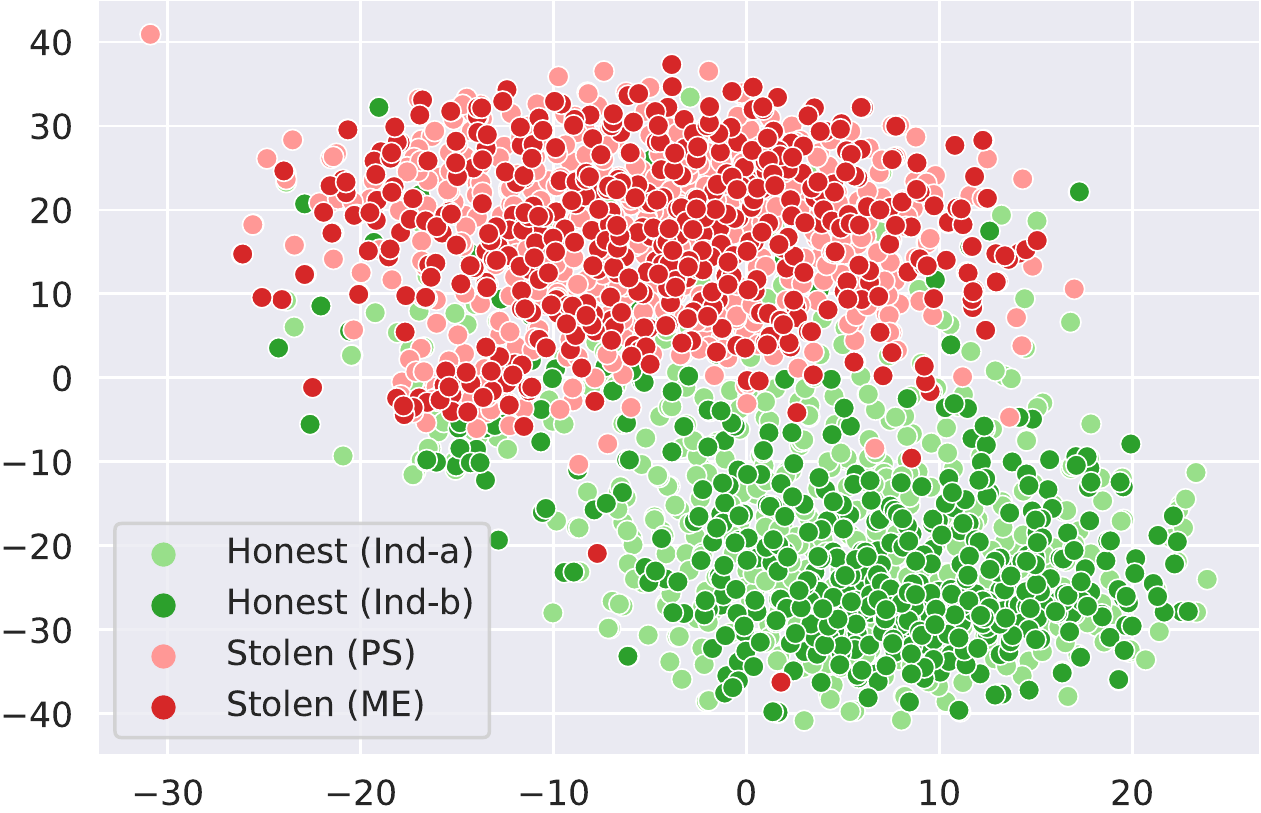}
	\caption{T-SNE visualization of characteristics learned by our method for stolen models and honest models.}
	\label{fig:t_sne}
\end{figure}

Figure~\ref{fig:t_sne} shows the T-SNE visualization of characteristics learned by our method. 
We plot the T-SNE figure by using outputs from the penultimate layer of the classifier and the dimension of the outputs is 2,048.
We clearly see that characteristics from stolen models including PS and ME are entangled together and have a clear boundary with that from honest models.

\subsection{Generations of Model Extraction Attacks}
\label{ssec:ana_generations}

\begin{figure}[!t]
	\centering
	\includegraphics[width=0.75\linewidth]{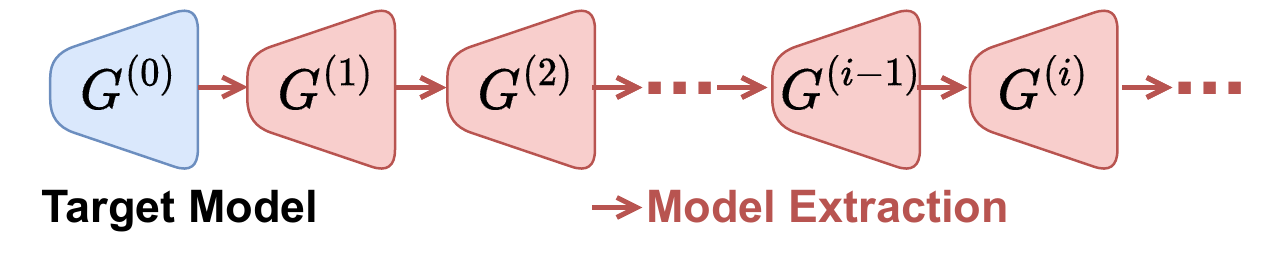}
	\caption{Generations of model extraction attacks.}
	\label{fig:ME_multi}
\end{figure}

Theoretically, model extraction attacks on GANs can continue forever like the process of biological heredity, as shown in Figure~\ref{fig:ME_multi}. 
Models produced
during this process, such as $G^{(i)}$, should be correctly identified by
ownership protection methods. 
This motivates us to investigate whether the protection performance will decrease with the number of generations of model extraction attacks. 
We emphasize this is our newly identified threat, which is not discussed in the literature about GAN ownership protection.

Here, we fix the number of generated samples as 1,000 and the target model is SNGAN trained on FFHQ-I. 
We mark the target model SNGAN as SNGAN$^{(0)}$, and the first generation of model extraction is marked as SNGAN$^{(1)}$, which means an adversary uses an attack model SNGAN to extract the target model SNGAN. 
We do not show the performance of the Ong method because it cannot defend against model extraction attacks.

\begin{figure}[!t]
	\centering
	\includegraphics[width=0.7\linewidth]{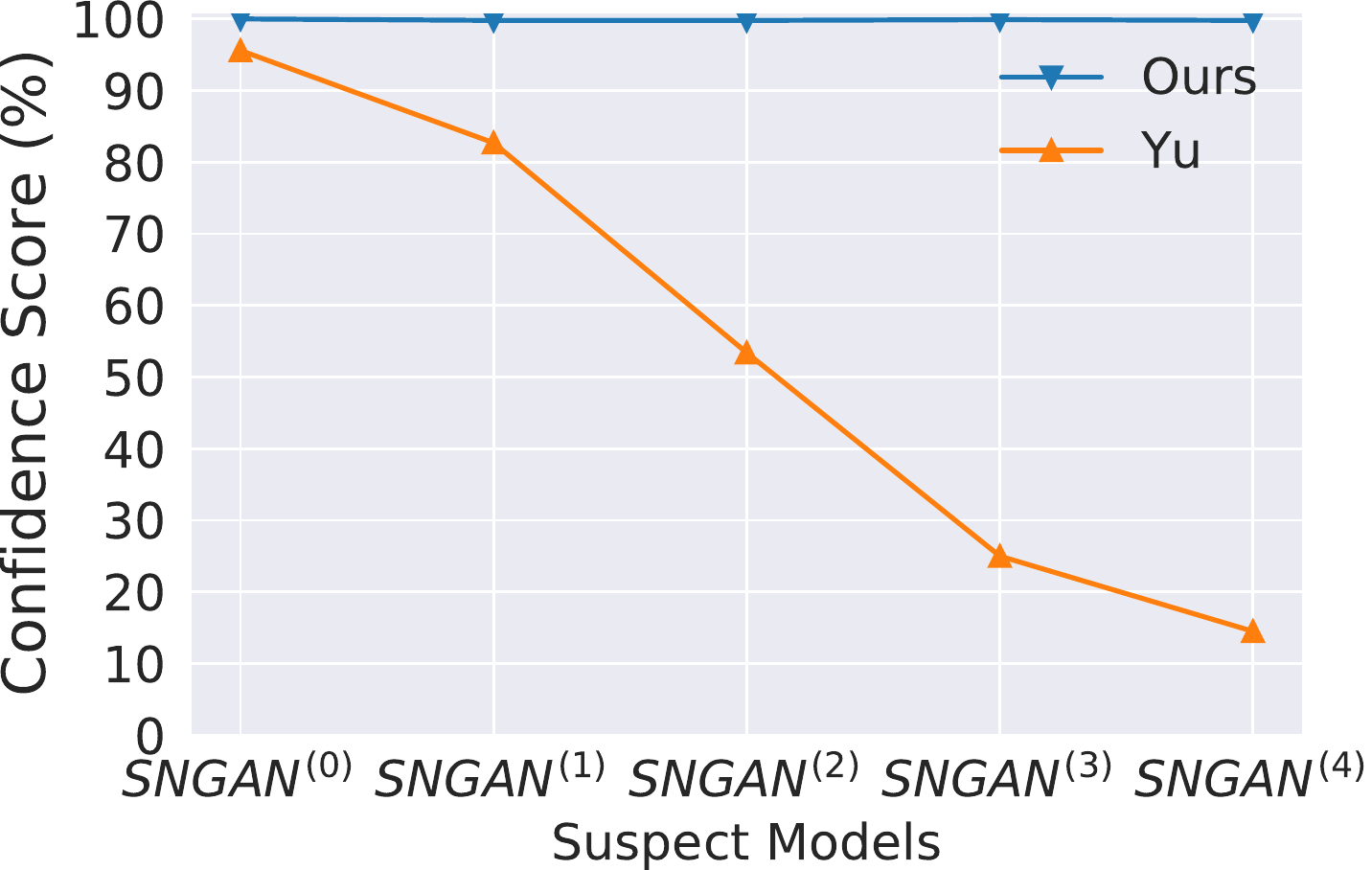}
	\caption{Protection performance with regard to the number of generations of model extraction attacks.}
	\label{fig:different_iter_ME}
\end{figure}

As shown in Figure~\ref{fig:different_iter_ME}, we can clearly observe that with the increase in the number of generations of model extraction, the Yu method becomes less and less confident. 
It also indicates that the fingerprint is not robust and more and more generated samples cannot extract the corresponding fingerprint. 
In contrast, our method still remains almost 100\% confident to verify ownership of the target model. 

\subsection{Number of Generated Samples}
\label{ssec:ana_queries}

\begin{figure}[!t]
	\centering
	\includegraphics[width=0.7\linewidth]{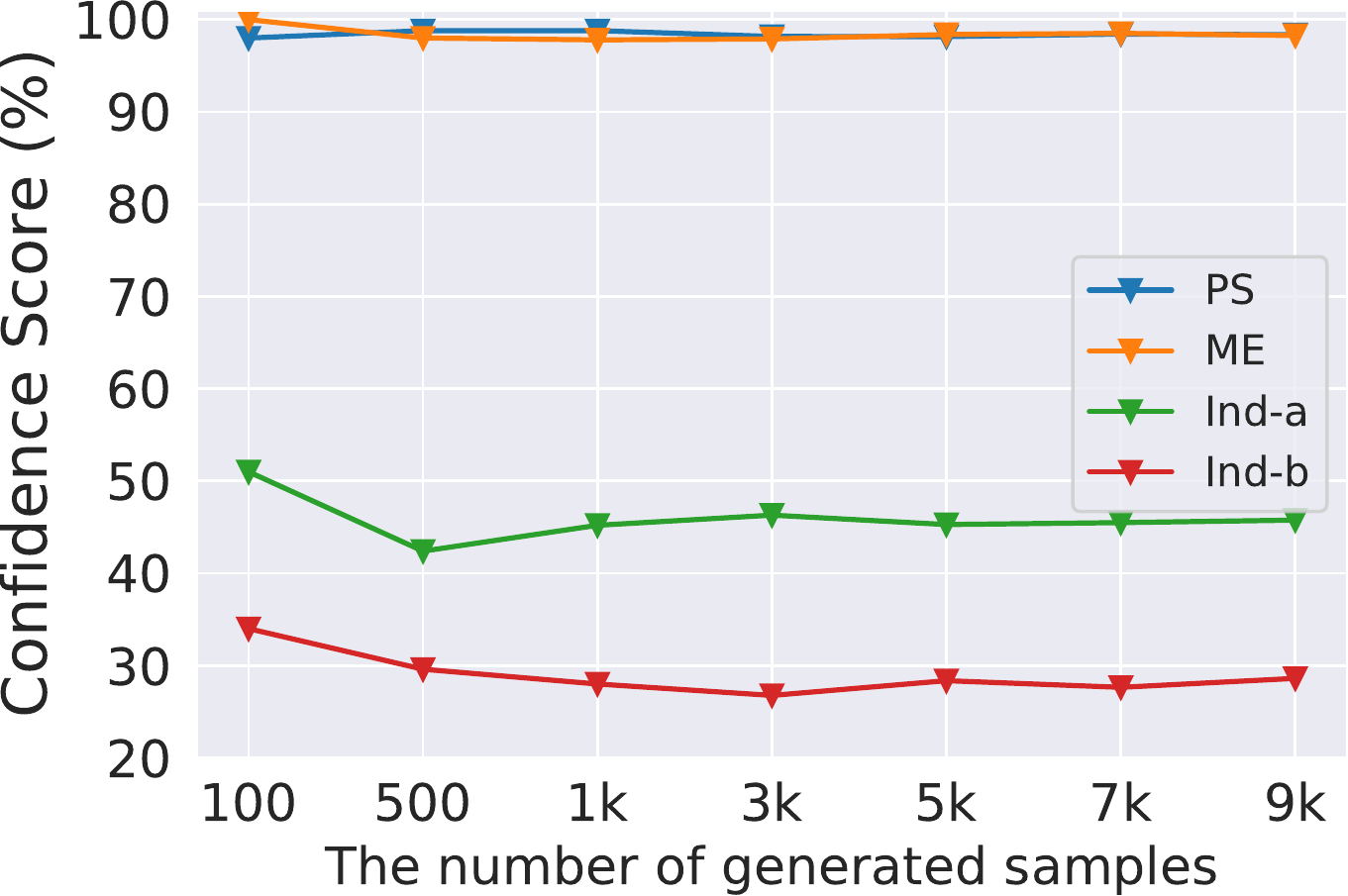}
	\caption{Protection performance with regard to the numbers of generated samples.}
	\label{fig:different_queries}
\end{figure}

Figure~\ref{fig:different_queries} presents the protection performance of our method under the different numbers of generated samples. 
The target model SNGAN is trained on FFHQ-I.
The ground truth of PS and ME is positive while that of Ind-a and Ind-b is negative.
We can clearly see that the confidence scores gradually remain stable after 1,000 generated samples on all suspect models.
It also shows that our protection method has advantages with respect to the \emph{efficiency}, i.e. it requires as few as 1,000 samples.

\subsection{Different Datasets}
\label{ssec:diff_dataset}
We now present the performance of our method on the Church dataset which is widely used in scene synthesis.
The detail of this dataset is also discussed in Section~\ref{ssec:Datasets}.
The target model is SNGAN trained on the Church-I dataset and achieves 12.96 FID.

\begin{figure*}[!t]
	\centering
	
	\subfigure[Verification performance.]{
		\includegraphics[width=0.48\columnwidth]{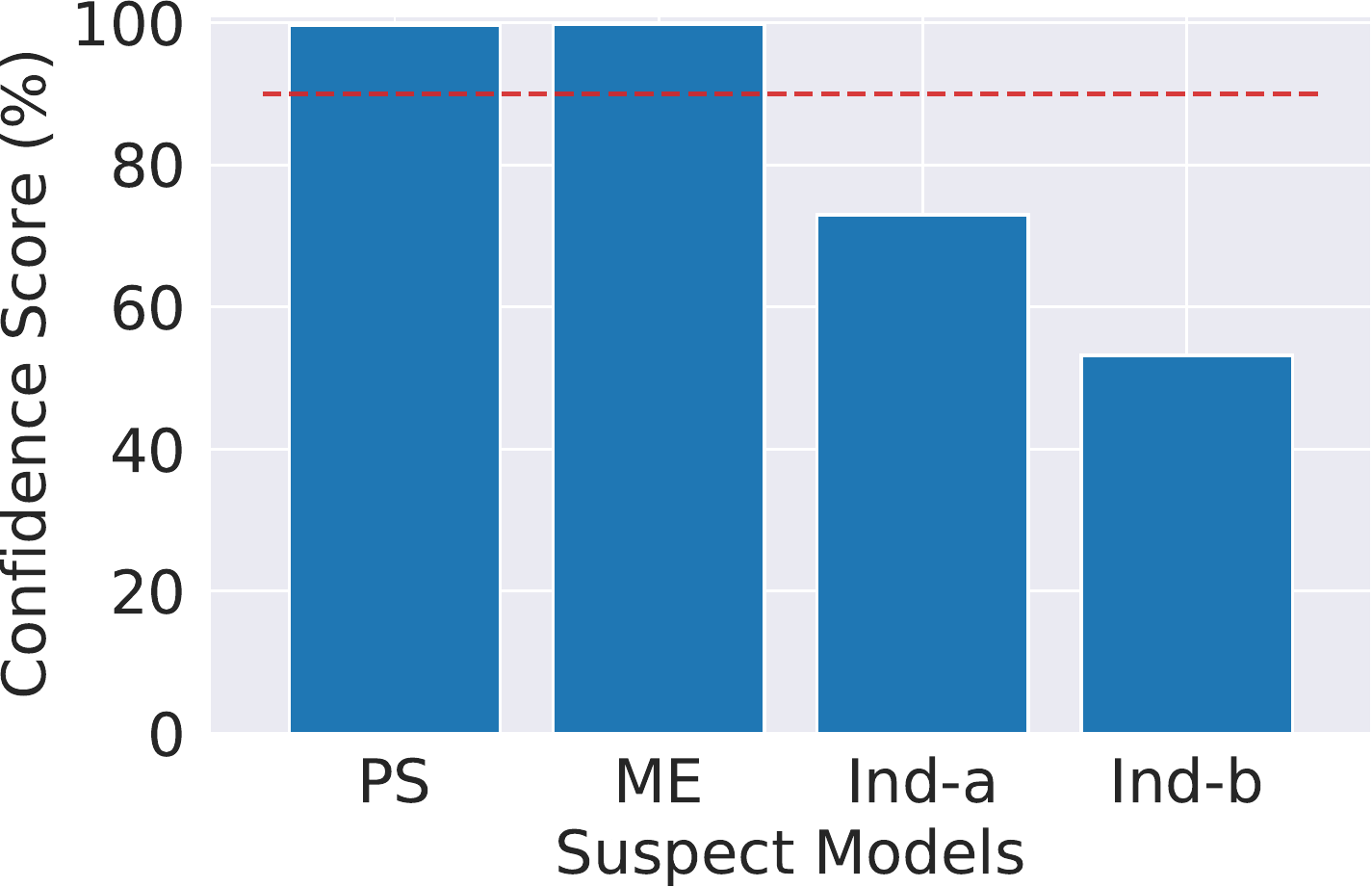}
		\label{fig:Eval_BA_sngan_church1}
	}
	\subfigure[Input perturbation.]{
	\includegraphics[width=0.48\columnwidth]{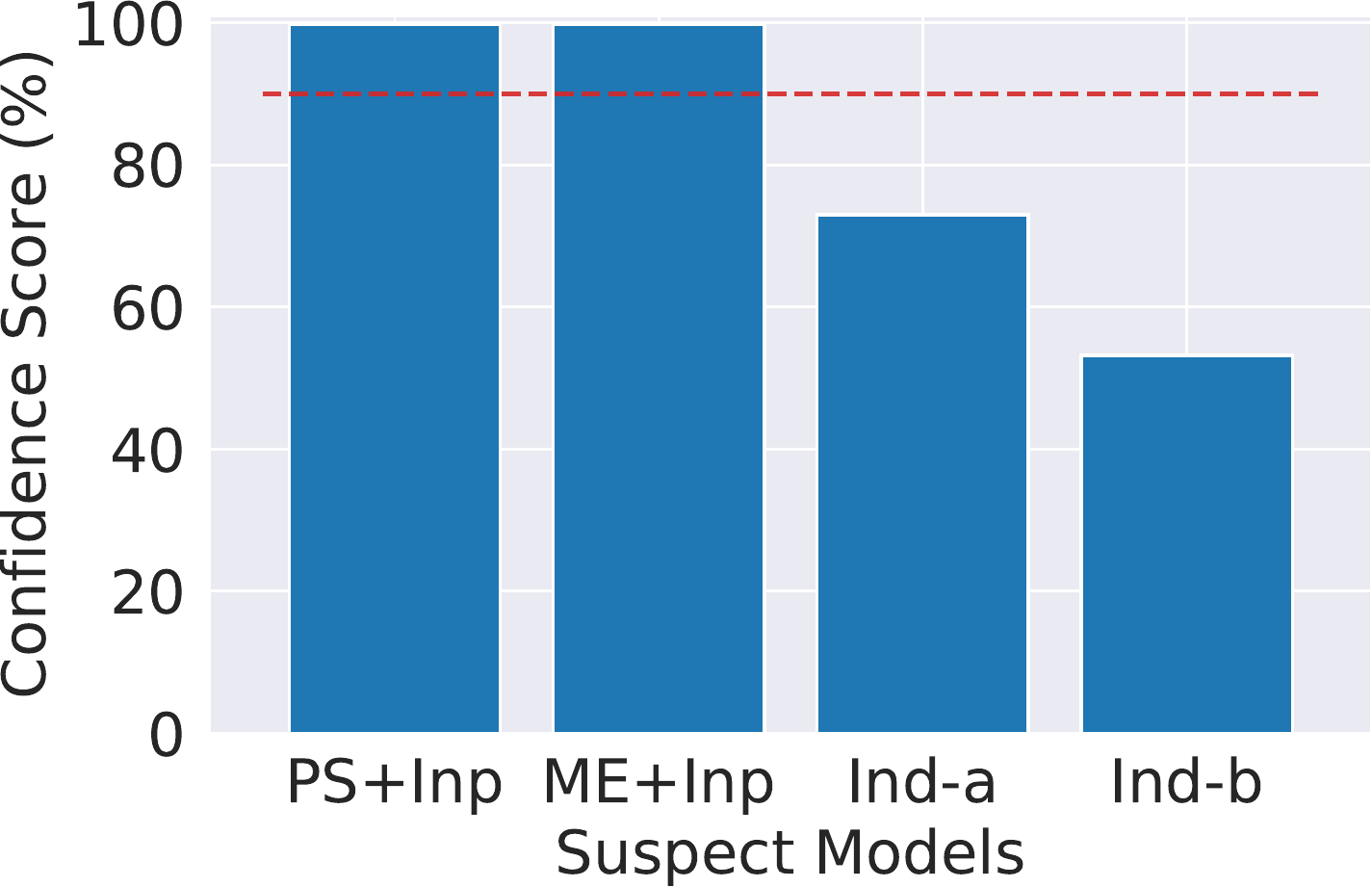}
	\label{fig:Eval_AdvA_sngan_church1_input}
}
	\subfigure[PS + output perturbation.]{
		\includegraphics[width=0.48\columnwidth]{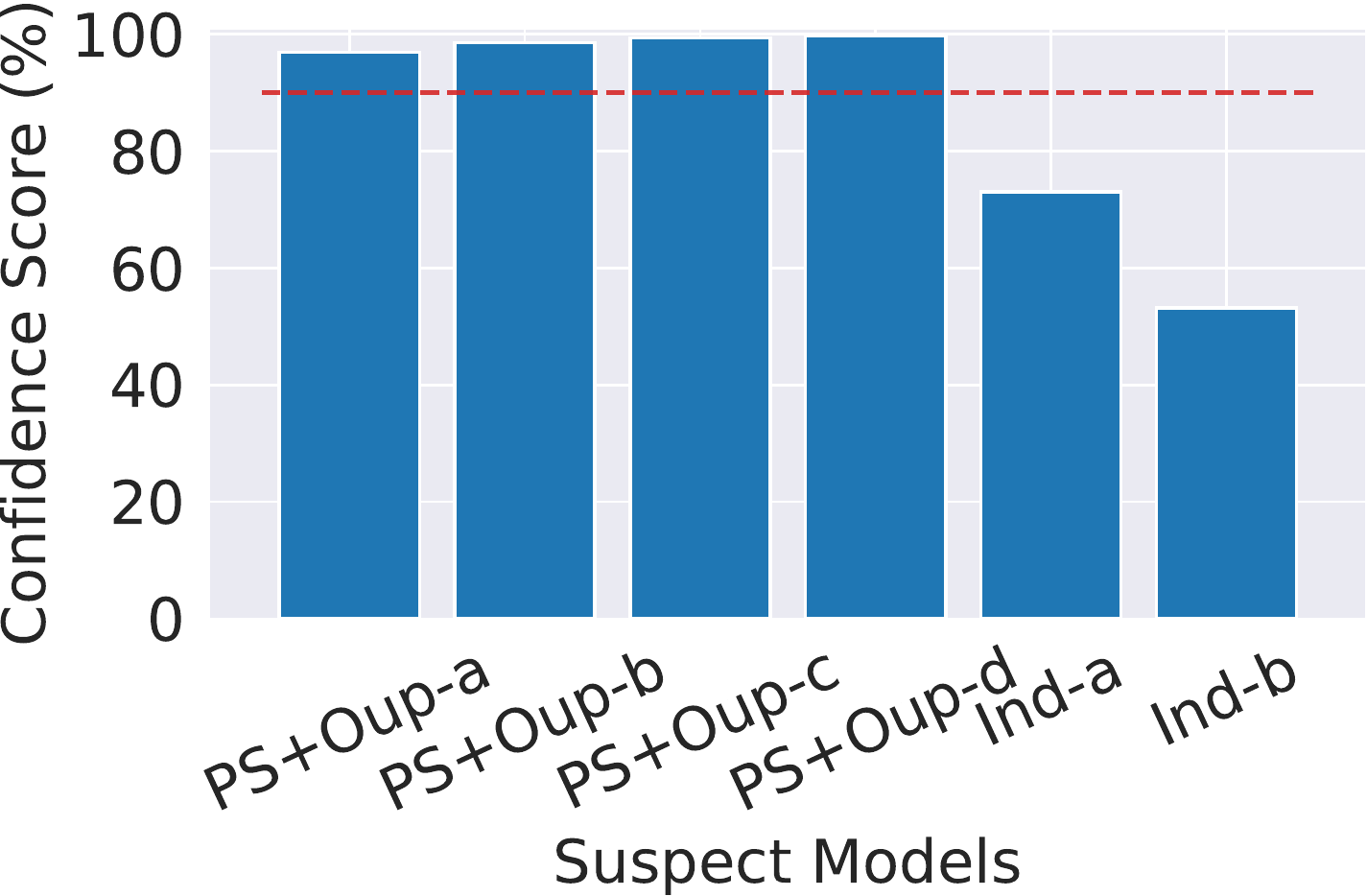}
		\label{fig:Eval_AdvA_sngan_church1_output_ps}
	}	
	\subfigure[ME + output perturbation.]{
		\includegraphics[width=0.48\columnwidth]{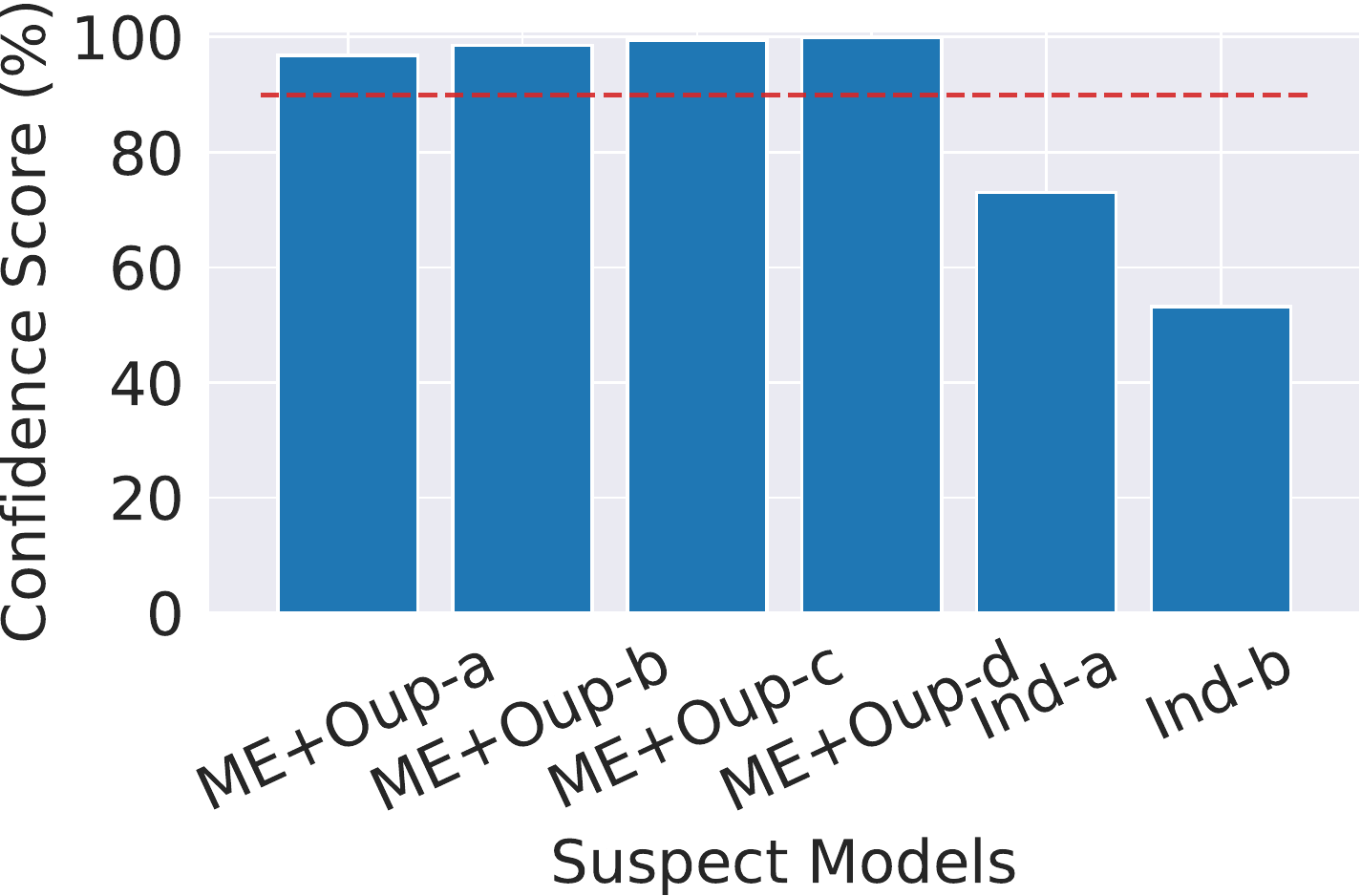}
		\label{fig:Eval_AdvA_sngan_church1_output_me}
	}
	\caption{Protection performance on target model SNGAN trained on Church-I. }
	\label{fig:Eval_oba_church}
\end{figure*}

Overall, our method on the Church dataset can achieve the same exceptional protection performance as that on the FFHQ dataset.
Figure~\ref{fig:Eval_BA_sngan_church1} shows verification performance on our method. 
We can clearly observe that our method can verify the positive suspect models as positive with high confidence.
As shown in Figure~\ref{fig:Eval_AdvA_sngan_church1_input}, Figure~\ref{fig:Eval_AdvA_sngan_church1_output_ps} and Figure~\ref{fig:Eval_AdvA_sngan_church1_output_me}, our method can still remain 100\% accuracy under the input perturbation and output perturbation.
Our method on fine-tuning attacks shows it fails to recognize positive suspect models as positive. We show details in Table~\ref{tab:Eval_AdvA_sngan_church1_finetuning} in Appendix, and 
we also analyze the performance on different target models in Section~\ref{sssec:diff_targets} in Appendix.

\section{Adaptive Attacks}
\label{sec:MPAdA}
Although researchers studying defense techniques strongly advocates that a new defense should be evaluated on adaptive attacks~\cite{tramer2020adaptive}, prior works on ownership protection on GANs do not adopt it.
In this work, for the first time, we present the performance of our method under adaptive attacks.
That is, we assume that adversaries have some knowledge of our protection method, and design a series of specific attacks to evade our method.

\begin{figure}[!t]
	\centering
	\includegraphics[width=0.70\linewidth]{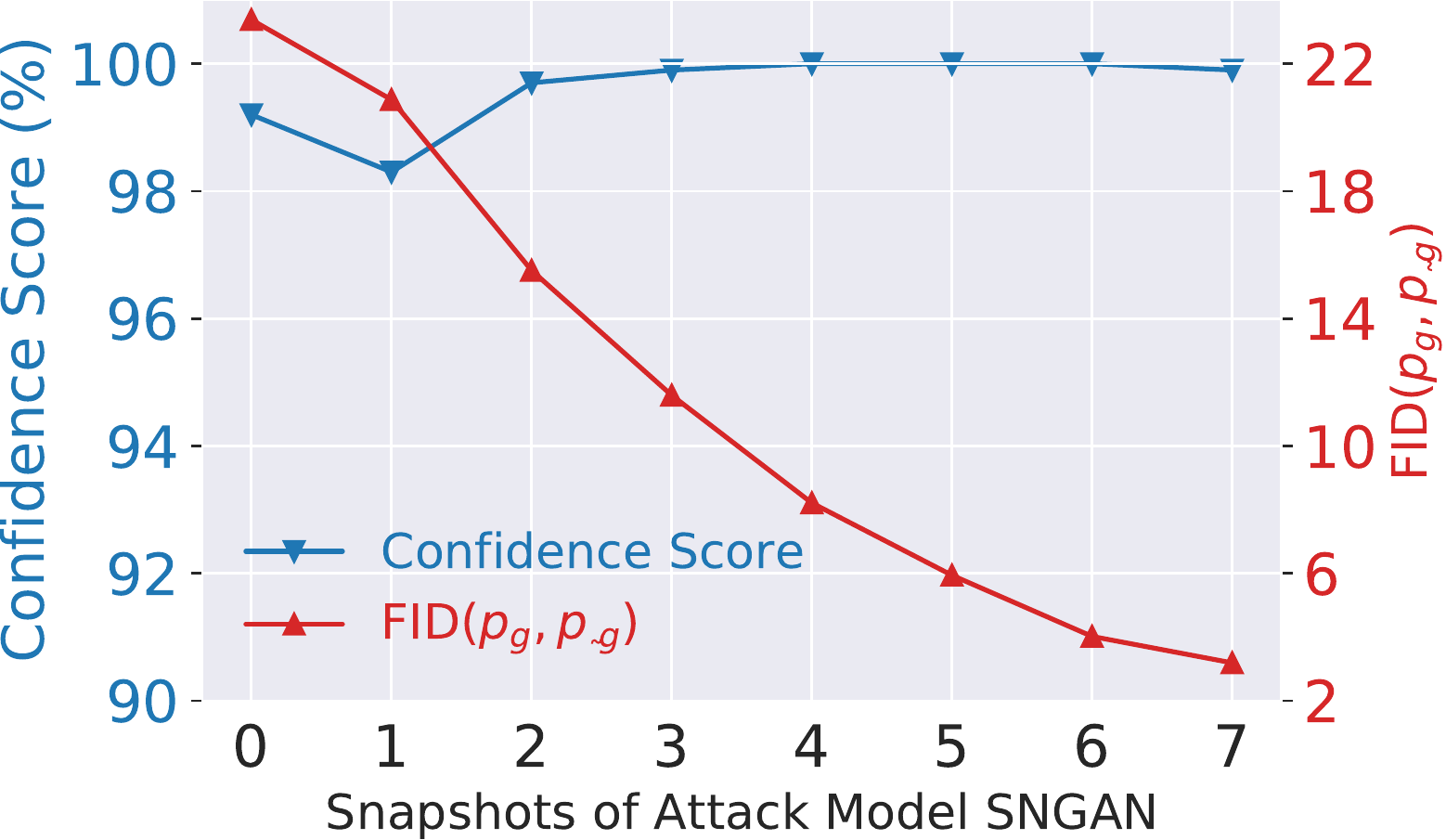}
	\caption{Protection performance under the adaptive attack~I. The target model SNGAN is trained on FFHQ-I.}
	\label{fig:adaptive1_sngan_sngan}
\end{figure}  

\begin{figure}[!t]
	\centering
	\includegraphics[width=0.70\linewidth]{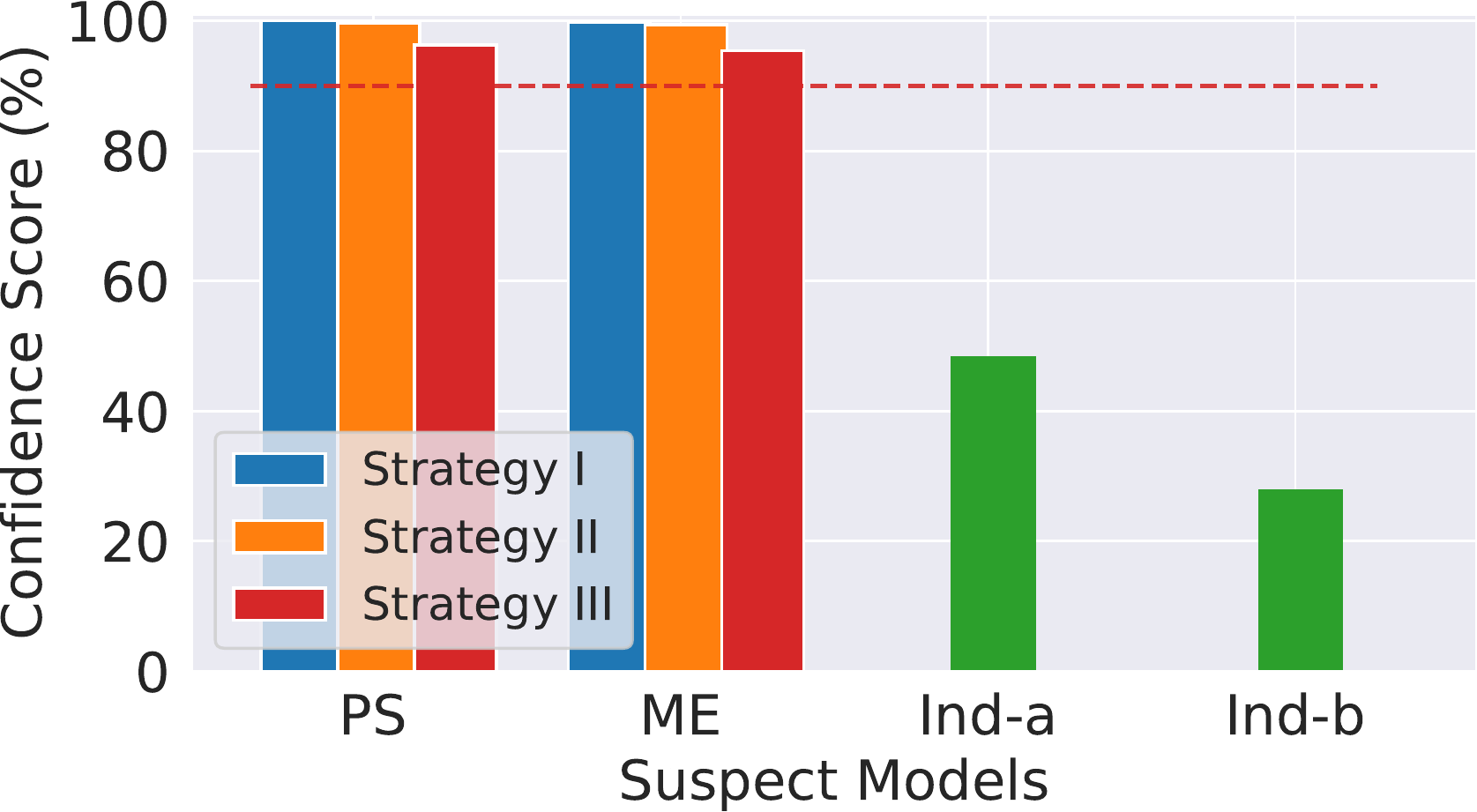}
	\caption{Protection performance under the adaptive attack~II. The target model SNGAN is trained on FFHQ-I.}
	\label{fig:adaptive2_sngan_sngan}
\end{figure}

We discuss two types of adaptive attack scenarios.
The main design principle is that we assume that adversaries perceive our method which is based on the common characteristics of a target model and its stolen models.
Therefore, the adversaries attempt to decrease the confidence score of our method by sacrificing model utility (i.e. the quality of generated images).
Specifically, in adaptive attack~I, adversaries choose an inferior performance GAN from multiple snapshots of a GAN when mounting model extraction attacks.
In adaptive attack~II, adversaries evade our verification by designing a series of output perturbations by choosing the magnitude of the perturbation. 

Figure~\ref{fig:adaptive1_sngan_sngan} shows protection performance under the adaptive attack~I.
Here, we choose an attack model SNGAN to extract the target model SNGAN trained on FFHQ-I.
We choose eight snapshots of SNGAN during model extraction attacks. The performance of the attack model SNGAN, i.e. FID($p_g, p_{\tilde g}$), ranges from 22 to 2, as depicted in the red line.
$p_g$ and $p_{\tilde g}$ are the implicit distribution of the target model and the attack model, respectively.
We can observe that confidence scores begin to decrease, then increase and remain at 100\%, with the decrease in FID of the attack model SNGAN.
In particular, the confidence scores of all snapshots are above 98\%, which indicates that our method can correctly recognize all snapshots as stolen models.

Figure~\ref{fig:adaptive2_sngan_sngan} shows protection performance under the adaptive attack~II.
Considering the model utility, we design three strategies (strategy~I, strategy~II and strategy~III) based on different magnitudes of four types of output perturbation.
Table~\ref{tab:adaptive2_sngan_sngan_parameters} summarizes the magnitudes of output perturbation of each strategy.
Note that we combine four types of output perturbation instead of single output perturbation.
In Figure~\ref{fig:ada_strategy_2} in Appendix, we visually present images generated by each strategy and the quality of generated images becomes much noisier and blurrier from strategy~I to strategy~III.
Overall, we can observe in Figure~\ref{fig:adaptive2_sngan_sngan} that our method still can recognize all positive suspect models, although the confidence score of each suspect model decreases from strategy~I to strategy~III.
In addition, although strategy~III can lower the confidence of our method, the model almost cannot be used due to the low quality of generated images visually. 
In practice, the loss in model utility can make the adversaries less competitive in market share, compared to legitimate model owners. 

\begin{table}
	\centering
	\caption{Magnitudes of output perturbation of the adaptive attack II. a: Additive Gaussian Noise; b: Gaussian Filtering; c: Gaussian Blurring; d: JPEG Compression.}
	\label{tab:adaptive2_sngan_sngan_parameters}
	\renewcommand{\arraystretch}{1.0}
\scalebox{0.90}{		
	
\begin{tabular}{ccccc} 
\specialrule{.15em}{.05em}{.05em}
	Strategy & \multicolumn{4}{c}{Output Perturbation}  \\ 
	\cline{2-5}
	& a     & b   & c   & d           \\ 
	\hline
	I        & 0.001 & 0.1 & 0.1 & 95          \\
	II       & 0.005 & 0.2 & 0.3 & 90          \\
	III      & 0.01  & 0.4 & 0.5 & 85          \\
\specialrule{.15em}{.05em}{.05em}
\end{tabular}
	
}
\end{table}

\section{Conclusion}
\label{sec:Conclusion}

In this paper, we have proposed a novel method to protect GAN ownership by leveraging the common characteristics of a target model and its stolen GANs.
Extensive experimental evaluations demonstrate that:
(a) In terms of model utility, our method can bring lossless fidelity, compared to models without protection, because it does not modify well-trained target models.
(b) In terms of robustness, our method can achieve new state-of-the-art protection performance, compared with watermark-based methods and fingerprint-based methods.
Furthermore, we have also shown that our method is still effective under two types of carefully designed adaptive attacks.
(c) In terms of undetectability, our method is undetectable for adversaries because it builds on a target model with normal training and does not rely on watermarks or fingerprints.
(d) In terms of efficiency,  our method requires about 1,000 generated samples to confidently verify the ownership of a GAN.
Finally, we also have performed a fine-grained analysis of our method from various aspects, such as visualizing learned characteristics, the stability of the performance with regard to the number of generations of model extraction attacks, the number of generated samples and different datasets.

Fine-tuning attacks remain a challenge for ownership protection on GANs. 
In future, we plan to design more powerful methods to defend against these types of attacks.
In addition, applying our protection method to other domains, such as table data synthesis and text generation, is also an interesting research direction.

\section*{Acknowledgments}
This research was funded in whole by the Luxembourg National Research Fund (FNR), grant reference 13550291.

{\small
\bibliographystyle{ieee_fullname}
\bibliography{reference}
}

\clearpage
\appendix
\section{Appendix}

\subsection{Details of Obfuscations}
\label{ssec:details_obfu}
In this subsection, we introduce obfuscation operations, including input perturbation, output perturbation, overwriting, and fine-tuning.

\smallskip\noindent
\textit{Input perturbation.} 
Given a trained GAN model~$G$, we can get a generated sample~$x_g$ from the GAN by a latent code~$z$ which is drawn from prior distribution~$P$, i.e., $x_g = G(z), z \sim P$. 
Input perturbation aims to modify the queries, i.e., latent codes. 
The reason why we consider this is that some works verify the ownership by specific latent codes $z' = T(z)$. $T$ is a function to transform a normal latent code $z$ to a specific latent code~$z'$.
Therefore, an adversary can perturb latent codes to evade this type of verification.
In this work, we adopt random input perturbation. 
Specifically, for any query, a target model resamples
latent codes from Gaussian distribution.

\smallskip\noindent
\textit{Output perturbation.} In addition to perturbing latent codes, an adversary can perturb generated samples. In this work, we consider the following four common operations.
\begin{enumerate}
	\vspace{0cm}\item[a.]\textit{Random Noise.} This operation adds noises into a generated sample~$x_g$. Common noises include Gaussian noise and Poisson noise. In this work, we choose Gaussian noise and the mean~$\mu$ and the variance~$\sigma$ control the strength of noises.
	\vspace{0cm}\item[b.]\textit{Filtering.} This operation aims to enhance some characteristics of an image. Common operations include mean filter, median filter and Gaussian filter.  In this work, we choose the Gaussian filter.
	\vspace{0cm}\item[c.]\textit{Blurring.} This operation makes a generated sample~$x_g$ less sharp by convolution. Common blurring operations include box blurring and Gaussian blurring. In this work, we choose Gaussian blurring.
	\vspace{0cm}\item[d.]\textit{Compression.} This operation is to compress the size of an image without significantly degrading the image quality. Common compression operations include lossless compression JPEG and lossy compression JPG. In this work, we choose JPEG compression.
\end{enumerate}

\smallskip\noindent
\textit{Overwriting.} 
This attack is to target a class of ownership protection methods utilizing watermarks or fingerprints. 
An adversary can encode a different watermark/fingerprint to overwrite the original watermark/fingerprint. 
Ideally, an ownership protection method should still verify the ownership in this case.
In this work, our proposed method does not rely on watermarks and fingerprints, thus intrinsically eliminating the threat of this attack.

\smallskip\noindent
\textit{Fine-tuning.} After obtaining a substitute model, an adversary may further fine-tune the model on their own dataset. Generally, an adversary can partially or wholly fine-tune the model.
Wholly fine-tuning refers that all weights of the model are fine-tuned while partially fine-tuning refers that the weights of some layers are frozen and the remaining are fine-tuned. In this work, we consider wholly fine-tuning in which we take the weights of the stolen model as initialization and retrain a GAN model on a new dataset FFHQ-II.

\subsection{Understanding In-depth}
\label{ssec:explan_ong}

\begin{figure}[!t]
	\centering
	\includegraphics[width=0.85\linewidth]{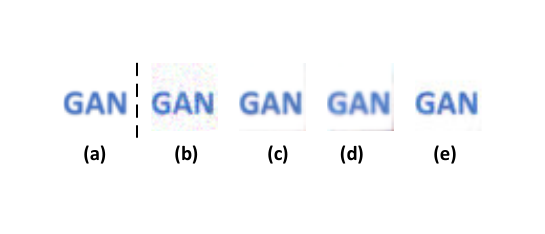}
	\caption{Watermarks under different output perturbations. (a) is the original watermark. From (b) to (e), the output perturbation operations are Additive Gaussian Noise, Gaussian Filtering, Gaussian Blurring, and JPEG Compression, respectively. The corresponding SSIM scores between (a) and each output perturbation are 84.085\%, 97.47\%, 99.34\%, 95.43\%, respectively.}
	\label{fig:wm_out_perturb}
\end{figure}

Figure~\ref{fig:wm_out_perturb} shows SSIM scores of watermarks under different output perturbations. 
Specifically, Figure~\ref{fig:wm_out_perturb} (a) is the watermark used in the Ong method.
Figure~\ref{fig:wm_out_perturb} (b) - Figure~\ref{fig:wm_out_perturb} (e) show the watermark under different output perturbations.
We can observe that only if the Ong method can extract watermarks, output perturbation attacks do not have a significant impact on the final decision.
This can explain the reason why the Ong Method on PS+output perturbation can perform well, as shown in Figure~\ref{fig:cmp_oup_ps}.

However, model extraction attacks and their derivative attacks (i.e. ME+obfuscations) make the Ong and Yu methods fail.
This is because these attacks severely undermine watermarks and fingerprints.
It also indicates that methods based on watermarks and fingerprints are not robust and too easily perturbed.

\subsection{Additional Results in Section 6}
\label{ssec:add_res_sec6}

\subsubsection{Performance on overwriting attacks}
\begin{figure}[!t]
	\centering
	\includegraphics[width=0.1\linewidth]{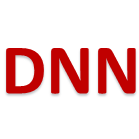}
	\caption{Watermarks used for overwriting attacks.}
	\label{fig:wm_overwriting}
\end{figure}

Figure~\ref{fig:wm_overwriting} shows a new watermark that is used for overwriting attacks for the Ong method. The original watermark is shown in Figure~\ref{fig:wm_out_perturb} (a).

Table~\ref{tab:Eval_AdvA_sngan_ffhq1_overwriting} presents protection performance under the overwriting attacks.
We do not show results for our method because our method does not rely on watermarks or fingerprints. 
Here, we only choose the PS model obtained by physical stealing because it can be perfectly recognized by both methods on physical stealing attacks. 
Intuitively, the protection method cannot verify this suspect PS model, which also means that other suspect models cannot be correctly verified.
We observe that the Ong and Yu methods cannot defend against this type of attack.

\begin{table}[]
	\centering
	\caption{Protection performance on  overwriting. The target model SNGAN is trained on FFHQ-I. The suspect model PS is the model obtained by physical stealing.
	Confi.: confidence; Pred.: prediction.}
	\label{tab:Eval_AdvA_sngan_ffhq1_overwriting}
	\renewcommand{\arraystretch}{1.0}
	\scalebox{0.85}{	
		
		\begin{tabular}{l|rr|rr}
			\specialrule{.15em}{.05em}{.05em}
			Types& \multicolumn{2}{c|}{Ong}  & \multicolumn{2}{c}{Yu}     \\
			   & Confi. Score(\%) & Pred. & Confi. Score(\%) & Pred.  \\ 
			\hline
			PS & 0.00             & 0      & 0.00             & 0      
			\\
			\specialrule{.15em}{.05em}{.05em}
		\end{tabular}
		
	}
\end{table}

\subsubsection{Performance on fine-tuning attacks}
We adopt wholly fine-tuning where all weights of each model are fine-tuned on FFHQ-II. Specifically, we take the weights of the stolen model as initialization and retrain a GAN model on FFHQ-II.

Table~\ref{tab:Eval_AdvA_sngan_ffhq1_finetuning*} reports protection performance under the fine-tuning attack. 
We observe that all protection methods cannot be robust against this attack.
The main reason is that fine-tuning enforces a GAN to learn the distribution of a new dataset. 
Due to the catastrophic forgetting of a neural network, previously embedded information, such 
as watermarks and fingerprints, cannot be kept. 
Similarly, the fine-tuned GAN that has learned a new distribution is no longer similar to the target model, which fails our method.

\begin{table}
	\centering
	\caption{Protection performance on fine-tuning. Target model SNGAN is trained on FFHQ-I.}
	\label{tab:Eval_AdvA_sngan_ffhq1_finetuning*}
	\renewcommand{\arraystretch}{1.0}
	\scalebox{0.75}{

	\begin{tabular}{l|rr|r|r|r|r} 
	\specialrule{.15em}{.05em}{.05em}
	Types
	& \multicolumn{2}{c|}{Ong}                                                                 & \multicolumn{2}{c|}{Yu}                                                                & \multicolumn{2}{c}{Ours}                                                               \\ 
	\hline
	& \multicolumn{1}{c|}{\begin{tabular}[c]{@{}c@{}}Confi. \\Score(\%)\end{tabular}} & Pred. & \multicolumn{1}{c|}{\begin{tabular}[c]{@{}c@{}}Confi.\\Score(\%)\end{tabular}} & Pred. & \multicolumn{1}{c|}{\begin{tabular}[c]{@{}c@{}}Confi.\\Score(\%)\end{tabular}} & Pred.  \\ 
	\hline
	PS & 0.00                                                                            & 0      & 0.00                                                                           & 0     & 40.10                                                                          & 0      \\ 
	\hline
	ME & 0.00                                                                            & 0      & 0.00                                                                           & 0     & 28.00                                                                          & 0      \\
	\specialrule{.15em}{.05em}{.05em}
\end{tabular}

	}
\end{table}

\subsection{Additional Results in Section 7}
\label{ssec:add_res_sec7}

\subsubsection{Performance on different datasets}
\label{sssec:diff_datsets}

Table~\ref{tab:Eval_AdvA_sngan_church1_finetuning} shows the protection performance of our method on fine-tuning attacks.
we take the weights of the stolen model as initialization and retrain a GAN model on Church-II.
Our evaluation shows that it fails to recognize positive suspect models as positive.

\begin{figure}[!t]
	\centering
	\includegraphics[width=0.7\linewidth]{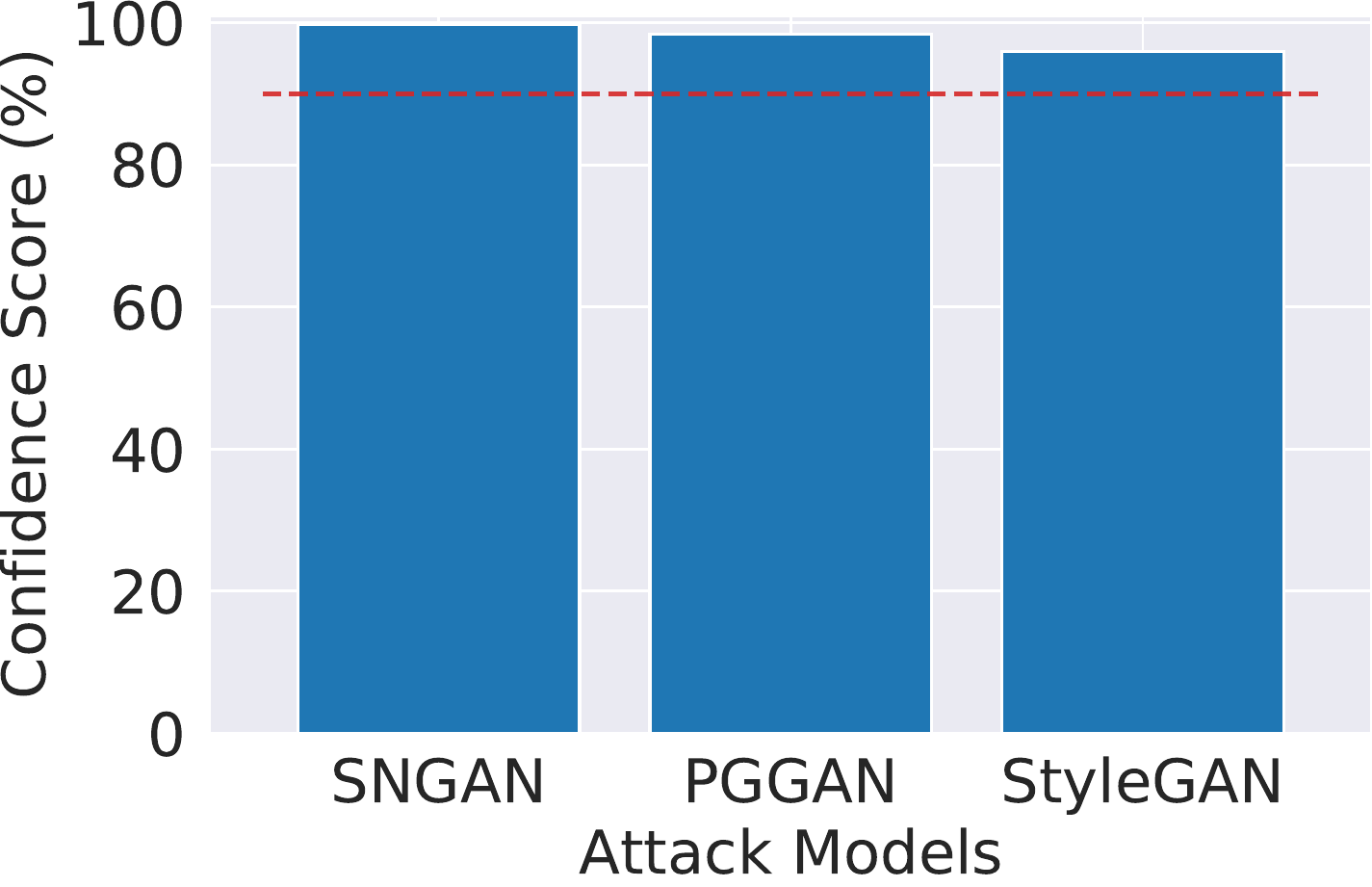}
	\caption{Robustness to more model extraction attacks. Protection performance under model extraction attacks with different GANs as attack models. Target model SNGAN is trained on Church-I. }
	\label{fig:Eval_More_ME_sngan_church1}
\end{figure}

Figure~\ref{fig:Eval_More_ME_sngan_church1} shows the results of our method in terms of robustness to different model extractions.
We observe that our method still performs perfectly regardless of the types of attack models.

\begin{table}
	\centering
	\caption{Protection performance on fine-tuning. Target model SNGAN is trained on Church-I.}
	\label{tab:Eval_AdvA_sngan_church1_finetuning}
	\renewcommand{\arraystretch}{1.0}
	\scalebox{0.85}{
		
	\begin{tabular}{c|c|r} 
	\specialrule{.15em}{.05em}{.05em}
	Types & Confi. Score(\%) & Pred.  \\ 
	\hline
	PS    & 50.50            & 0      \\ 
	\hline
	ME    & 46.40            & 0      \\
	\specialrule{.15em}{.05em}{.05em}
\end{tabular}
		
	}
\end{table}

\subsubsection{Performance on different target models}
\label{sssec:diff_targets}

We also show the protection performance of our method on the target model StyleGAN.
The StyleGAN is trained on FFHQ-I and achieves 8.76 FID.

\begin{figure*}[!t]
	\centering
	
	\subfigure[Verification performance.]{
		\includegraphics[width=0.47\columnwidth]{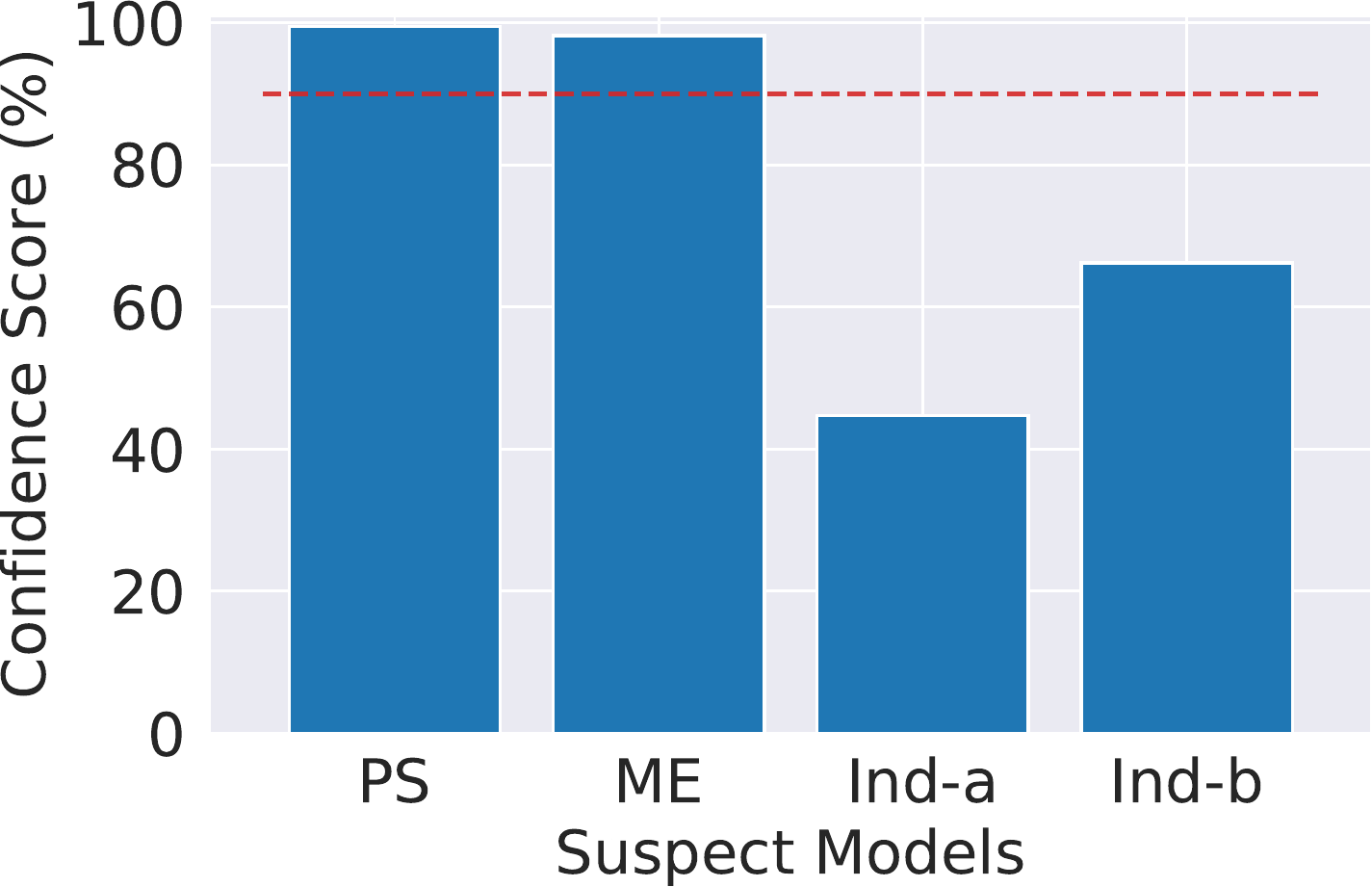}
		\label{fig:Eval_BA_stylegan_ffhq1}
	}
	\subfigure[Input perturbation.]{
		\includegraphics[width=0.47\columnwidth]{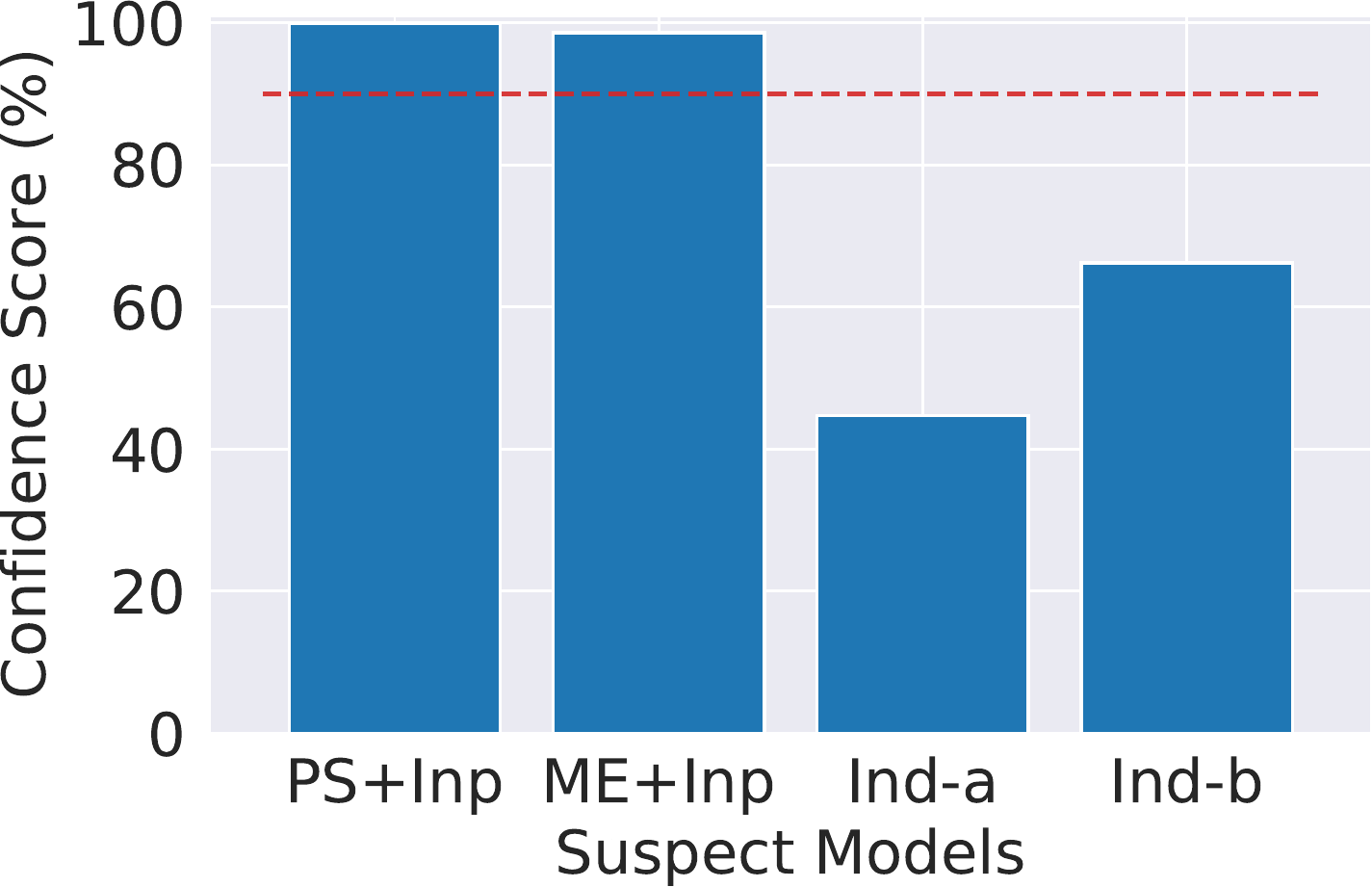}
		\label{fig:Eval_AdvA_stylegan_ffhq1_input}
	}
	\subfigure[PS + output perturbation.]{
		\includegraphics[width=0.47\columnwidth]{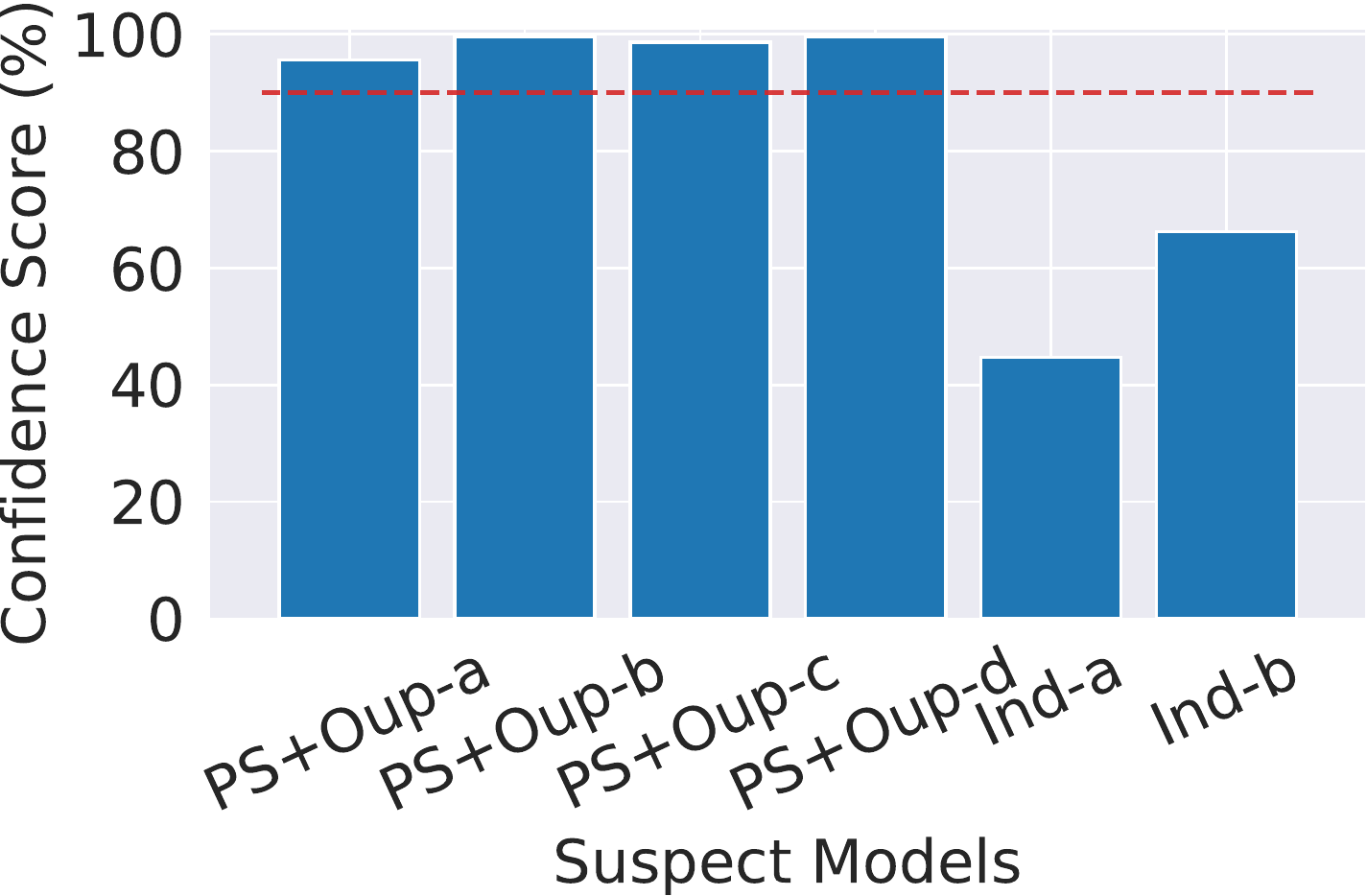}
		\label{fig:Eval_AdaA_stylegan_ffhq1_output_ps}
	}	
	\subfigure[ME + output perturbation.]{
		\includegraphics[width=0.47\columnwidth]{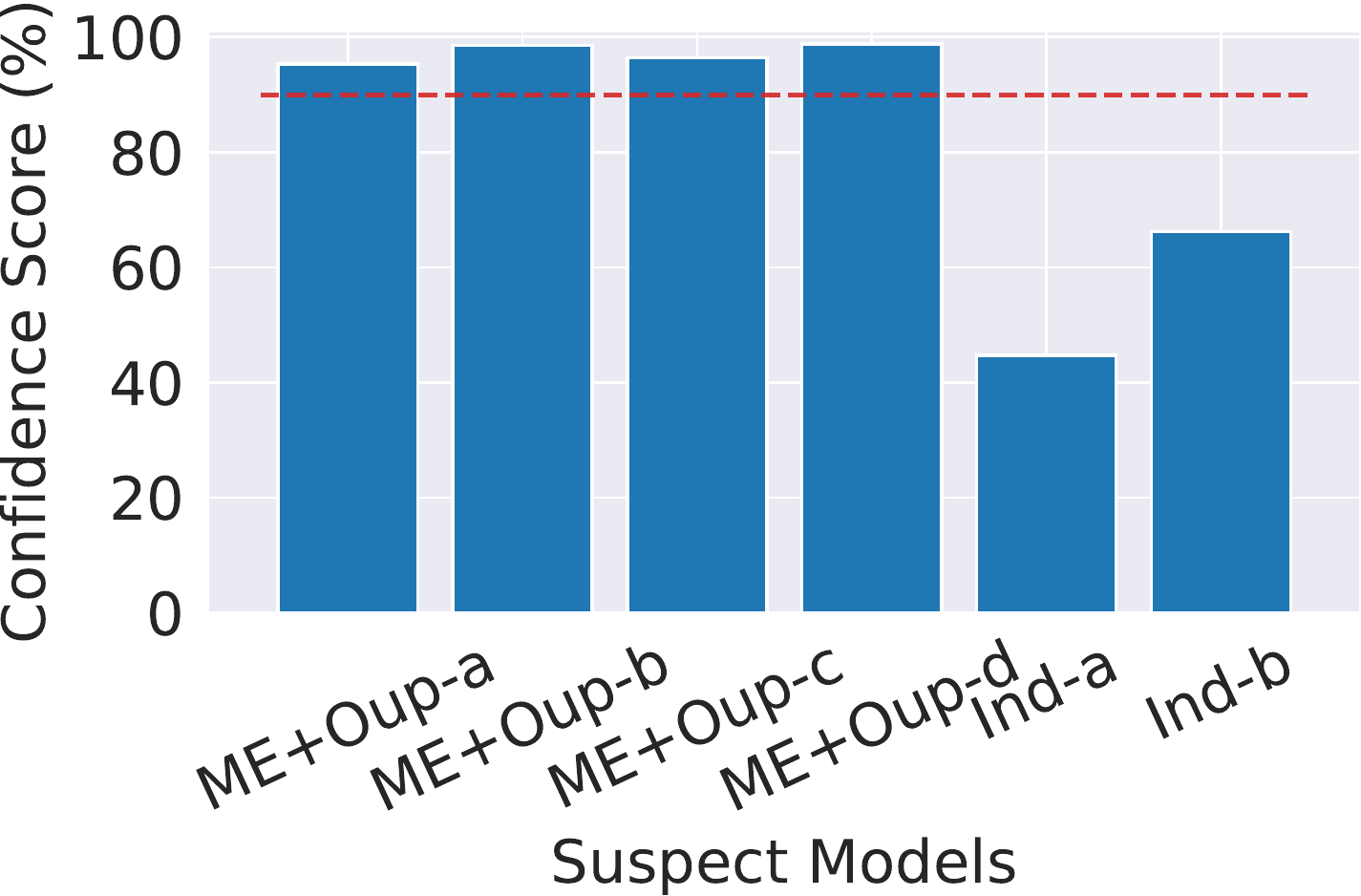}
		\label{fig:Eval_AdaA_stylegan_ffhq1_output_me}
	}
	\caption{Protection performance on target model StyleGAN trained on FFHQ-I. }
	\label{fig:Eval_oba_stylegan}
\end{figure*}

\begin{figure}[!t]
	\centering
	\includegraphics[width=0.7\linewidth]{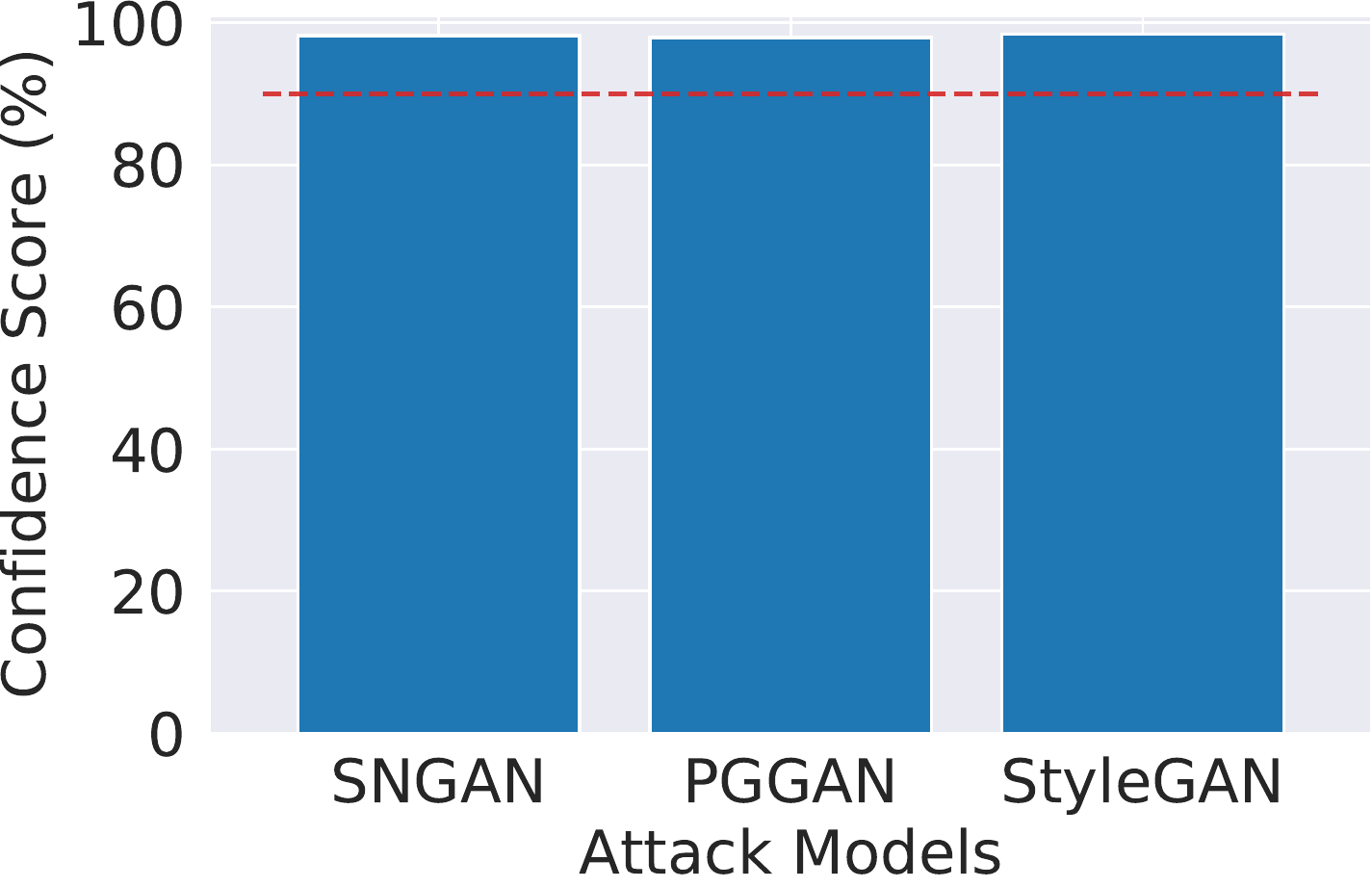}
	\caption{Robustness to more model extraction. Protection performance under model extraction with different GANs as attack models. Target model StyleGAN is trained on FFHQ-I. }
	\label{fig:Eval_More_ME_stylegan_ffhq1}
\end{figure}

Overall, our method still has competitive protection performance on the target model StyleGAN. 
Our method can achieve 100\% accuracy on verification performance for all suspect models, as shown in Figure~\ref{fig:Eval_BA_stylegan_ffhq1}.
In terms of obfuscations, 100\% accuracy can be seen on input perturbation attacks and four types of output perturbation attacks, as depicted in Figure~\ref{fig:Eval_AdvA_stylegan_ffhq1_input}, Figure~\ref{fig:Eval_AdaA_stylegan_ffhq1_output_ps} and  Figure~\ref{fig:Eval_AdaA_stylegan_ffhq1_output_me} respectively.
In the face of fine-tuning attacks, we can see in Table~\ref{tab:Eval_AdvA_stylegan_ffhq1_finetuning} that our method still fails to recognize these positive suspect models.
Figure~\ref{fig:Eval_More_ME_stylegan_ffhq1} shows the results of our method in terms of robustness to different model extractions.
We observe that our method still performs perfectly regardless of the types of attack models.

\begin{table}
	\centering
	\caption{Protection performance on fine-tuning. Target model StyleGAN is trained on FFHQ-I.}
	\label{tab:Eval_AdvA_stylegan_ffhq1_finetuning}
	\renewcommand{\arraystretch}{1.0}
	\scalebox{0.85}{
		
\begin{tabular}{c|c|r} 
	\specialrule{.15em}{.05em}{.05em}
	Types & Confi. Score(\%) & Pred.  \\ 
	\hline
	PS    & 35.90            & 0      \\ 
	\hline
	ME    & 49.40            & 0      \\
	\specialrule{.15em}{.05em}{.05em}
\end{tabular}		
		
	}	
\end{table}

\subsection{Image Quality under Adaptive Attack II}
\begin{figure}[!t]
	\centering
	\subfigure[No perturbation.]{
		\includegraphics[width=0.95\linewidth]{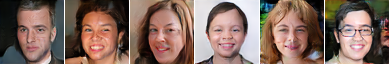}
		\label{fig:ada_strategy_no}
	}	
	\subfigure[Strategy I.]{
		\includegraphics[width=0.95\linewidth]{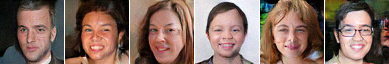}
		\label{fig:ada_strategy_I}
	}
	\subfigure[Strategy II.]{
		\includegraphics[width=0.95\linewidth]{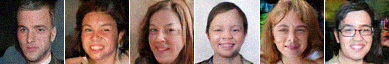}
		\label{fig:ada_strategy_II}
	}
	\subfigure[Strategy III.]{
		\includegraphics[width=0.95\linewidth]{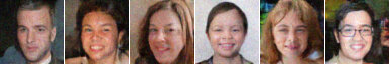}
		\label{fig:ada_strategy_III}
	}
	\caption{Adaptive attack II. From (a) strategy I to (c) strategy III, the magnitude of output perturbation gradually increases. The corresponding magnitudes are shown in Table~\ref{tab:adaptive2_sngan_sngan_parameters}. The average SSIM score for strategy I, strategy II, and strategy III is 92.20\%, 82.97\%, and 82.10\%, respectively.}
	\label{fig:ada_strategy_2}
\end{figure}

Figure~\ref{fig:ada_strategy_2} shows the image quality of three strategies under adaptive attack II.
Adversaries need to trade off model utility, i.e. the quality of generated images, and copyright infringement risks. 
Although a large magnitude output perturbation can help adversaries evade ownership detection, it will lower model utility. 
It also makes the stolen models less competitive, compared to the target models.

\subsection{Suspect Models}
\label{ssec:Suspect_Models}

\subsubsection{Implementation details of suspect models}
\label{ssec:imp_Suspect_Models}
For positive suspect models, models from physical stealing (marked as PS) are the same as target models.
We use model extraction attacks proposed in the work~\cite{hu2021model} to construct models from model extraction (marked as ME).
Specifically, given $m$ generated samples from a target model~$G_{\it tar}$, the adversaries retrain a model (also called the substitute model or attack model) on $m$ generated samples. Attack models can use any architecture, such as SNGAN, PGGAN, or StyleGAN. 
We study protection performance under model extraction attacks with different GANs as attack models in Section~\ref{ssec:Gen_ME}. 

For negative suspect models, Ind-a is trained with the same
architectures of target models but on dataset II.
Ind-b is trained with the same architectures and datasets of target models but
uses different seeds, i.e. different random initializations.
Using different seeds means models trained on different training environments and optimization processes. 
Theoretically, models trained with different seeds should be different, because they do not derive from model extraction and physical stealing, and they are honest models with independent training.
Thus, an ownership protection method should be able to differentiate them.
Here, setups for negative suspect models are very similar to those for target models because we aim to test whether a protection method hurt honest model providers in the strong assumption setting. 
This requires that a protection method should be extremely robust.

\begin{table*}
	\centering
	\caption{Performance of suspect positive models. The target model is SNGAN trained on FFHQ-I. It is corresponding to Figure~\ref{fig:cmp_basic}.}
	\label{tab:perf_pos_susp_sngan_ffhq}
	\renewcommand{\arraystretch}{1.3}
	\scalebox{1.0}{

\begin{tabular}{|l|r|r|r|r|r|r|} 
	\hline
	Types & \multicolumn{2}{c|}{Ong}                                & \multicolumn{2}{c|}{Yu}                                 & \multicolumn{2}{c|}{Ours}                                \\ 
	\hline
	& \multicolumn{2}{l|}{Attack Performance}                 & \multicolumn{2}{l|}{Attack Performance}                 & \multicolumn{2}{l|}{Attack Performance}                  \\ 
	\hline
	& \multicolumn{1}{l|}{FID($p_{\tilde g}, p_r$)} & \multicolumn{1}{l|}{FID($p_{\tilde g}, p_g$)} & \multicolumn{1}{l|}{FID($p_{\tilde g}, p_r$)} & \multicolumn{1}{l|}{FID($p_{\tilde g}, p_g$)} & \multicolumn{1}{l|}{FID($p_{\tilde g}, p_r$)} & \multicolumn{1}{l|}{FID($p_{\tilde g}, p_g$)}  \\ 
	\hline
	PS    & 20.14                      & 0.00                       & 26.46                      & 0.00                       & 20.25                      & 0.00                        \\ 
	\hline
	ME    & 25.47                      & 2.52                       & 30.83                      & 3.23                       & 27.60                      & 3.13                        \\
	\hline
\end{tabular}

}
\end{table*}

\subsubsection{Performance of suspect models}
\label{ssec:Perfor_Suspect_Models}

In this section, we show the performance of suspect models.
Overall, we choose suspect models with the best performance in each setting.
For positive suspect models, that is these models which are obtained by physical stealing or model extraction attacks, we use FID($p_{\tilde g}, p_r$) and FID($p_{\tilde g}, p_g$) to demonstrate the performance.
Here, $p_{\tilde g}$ is the implicit distribution of a suspect model or an attack model. $p_r$ is the implicit distribution of a training set. $p_g$ is the implicit distribution of a target model.
FID($p_{\tilde g}, p_g$) represents the similarity between an attack model and a target model, while FID($p_{\tilde g}, p_r$) represents the similarity between an attack model and the training set of a target model.
In the work~\cite{hu2021model}, they are also called fidelity and accuracy, respectively.
Here, we explicitly use FID($p_{\tilde g}, p_r$) and FID($p_{\tilde g}, p_g$) to report the performance of suspect positive models.
For negative suspect models, we use FID($p_r, p_g$) for evaluation.

Table~\ref{tab:perf_pos_susp_sngan_ffhq} and Table~\ref{tab:perf_neg_susp_sngan_ffhq} show positive and negative suspect models for target model SNGAN trained on FFHQ.
Table~\ref{tab:perf_cmp_More_ME} shows the performance of suspect positive models constructed by model extraction with different GANs.
Table~\ref{tab:perf_pos_susp_sngan_church} and Table~\ref{tab:perf_neg_susp_sngan_church} show positive and negative suspect models for target model SNGAN trained on Church.
Table~\ref{tab:perf_pos_susp_stylegan_ffhq} and Table~\ref{tab:perf_neg_susp_stylegan_ffhq} show positive and negative suspect models for target model StyleGAN trained on FFHQ.

\begin{table}
	\centering
	\caption{Performance of suspect negative models. The target model is SNGAN trained on FFHQ-I. FID($p_r, p_g$) is used for evaluation. }
	\label{tab:perf_neg_susp_sngan_ffhq}
	\renewcommand{\arraystretch}{1.3}
	\scalebox{1.0}{

\begin{tabular}{|l|r|r|r|} 
	\hline
	Types & \multicolumn{1}{c|}{Ong} & \multicolumn{1}{c|}{Yu} & \multicolumn{1}{c|}{Ours}  \\ 
	\hline
	Ind-a & 17.84                    & 17.84                   & 17.84                      \\ 
	\hline
	Ind-b & 20.63                    & 29.13                   & 17.15                      \\
	\hline
\end{tabular}

}
\end{table}

\begin{table}
	\centering
	\caption{Performance of suspect positive models. The target model is SNGAN trained on FFHQ-I. It is corresponding to Figure~\ref{fig:cmp_More_ME}.}
	\label{tab:perf_cmp_More_ME}
	\renewcommand{\arraystretch}{1.3}
	\scalebox{0.75}{
		
		\begin{tabular}{|l|r|r|l|l|} 
			\hline
			Attack Models & \multicolumn{2}{c|}{Ong}                & \multicolumn{2}{c|}{Yu}                  \\ 
			\hline
			& \multicolumn{2}{c|}{Attack Performance} & \multicolumn{2}{c|}{Attack Performance}  \\ 
			\hline
			& FID($p_{\tilde g}, p_r$) & FID($p_{\tilde g}, p_g$)                           & FID($p_{\tilde g}, p_r$) & FID($p_{\tilde g}, p_g$)                           \\ 
			\hline
			SNGAN         & 25.47 & 2.52                            & 30.83 & 3.23                             \\ 
			\hline
			PGGAN         & 23.58 & 2.33                            & 29.29 & 1.80                             \\ 
			\hline
			StyleGAN      & 21.62 & 2.70                            & 27.27 & 2.62                             \\ 
			\hline
			& \multicolumn{2}{c|}{Ours}               &       &                                  \\ 
			\hline
			SNGAN         & 27.60 & 3.13                            &       &                                  \\ 
			\hline
			PGGAN         & 23.11 & 2.56                            &       &                                  \\ 
			\hline
			StyleGAN      & 21.07 & 2.97                            &       &                                  \\
			\hline
		\end{tabular}
		
	}

\end{table}

\begin{table}
\centering
\caption{Performance of suspect positive models. The target model is SNGAN trained on Church-I. It is corresponding to Figure~\ref{fig:Eval_oba_church}.}
\label{tab:perf_pos_susp_sngan_church}
\renewcommand{\arraystretch}{1.3}
\scalebox{1.0}{

\begin{tabular}{|l|r|r|} 
	\hline
	Types & \multicolumn{2}{l|}{Attack Performance}  \\ 
	\hline
	& FID($p_{\tilde g}, p_r$) & FID($p_{\tilde g}, p_g$)                            \\ 
	\hline
	PS    & 12.96 & 0.00                             \\ 
	\hline
	ME    & 19.17 & 3.64                             \\
	\hline
\end{tabular}

}
\end{table}

\begin{table}
	\centering
	\caption{Performance of suspect negative models. The target model is SNGAN trained on Church-I.}
	\label{tab:perf_neg_susp_sngan_church}
\renewcommand{\arraystretch}{1.3}
\scalebox{1.0}{

\begin{tabular}{|l|r|} 
	\hline
	Types & \multicolumn{1}{c|}{FID($p_r, p_g$)}  \\ 
	\hline
	Ind-a & 13.35                     \\ 
	\hline
	Ind-b & 14.56                     \\
	\hline
\end{tabular}

}
\end{table}

\begin{table}
	\centering
	\caption{Performance of suspect positive models. The target model is StyleGAN trained on FFHQ-I. It is corresponding to Figure~\ref{fig:Eval_oba_stylegan}.}
	\label{tab:perf_pos_susp_stylegan_ffhq}
	\renewcommand{\arraystretch}{1.3}
	\scalebox{1.0}{

\begin{tabular}{|l|r|r|} 
	\hline
	Types & \multicolumn{2}{l|}{Attack Performance}  \\ 
	\hline
	& FID($p_{\tilde g}, p_r$) & FID($p_{\tilde g}, p_g$)                              \\ 
	\hline
	PS    & 8.76  & 0.00                             \\ 
	\hline
	ME    & 12.61 & 4.12                             \\
	\hline
\end{tabular}

	}
\end{table}

\begin{table}
	\centering
	\caption{Performance of suspect negative models. The target model is StyleGAN trained on FFHQ-I.}
	\label{tab:perf_neg_susp_stylegan_ffhq}
	\renewcommand{\arraystretch}{1.3}
	\scalebox{1}{
		
\begin{tabular}{|l|r|} 
	\hline
	Types & \multicolumn{1}{c|}{FID($p_r, p_g$)}  \\ 
	\hline
	Ind-a & 17.84                      \\ 
	\hline
	Ind-b & 8.92                       \\
	\hline
\end{tabular}

	}

\end{table}

\end{document}